\documentclass[usegraphicx,usenatbib]{mn2e}
\pdfoutput=1
\usepackage{graphicx}
\usepackage{dcolumn}
\usepackage{myaasmacros}
\usepackage{mathrsfs}
\usepackage{amsmath}
\usepackage{array,multirow}
\usepackage{amssymb}
\usepackage{amsfonts}
\usepackage{graphicx}
\usepackage{wasysym}
\usepackage{wrapfig}
\usepackage{dblfloatfix}
\usepackage{hyperref}
\usepackage{booktabs}
\usepackage{float}

\usepackage{pgfplots} 
\pgfplotsset{compat=newest} 
\pgfplotsset{plot coordinates/math parser=false} 
\newlength\figureheight 
\newlength\figurewidth 

\newcommand{\Ramses}{{\sc ramses}}
\newcommand{\RamsesRT}{{\sc ramses-rt}}

\newcommand{\maxl}{{{max43}}}
\newcommand{\minl}{{{min43}}}
\newcommand{\verl}{{{ver43}}}
\newcommand{\qsol}{{{qso43}}}
\newcommand{\maxh}{{{max46}}}
\newcommand{\minh}{{{min46}}}
\newcommand{\verh}{{{ver46}}}
\newcommand{\qsoh}{{{qso46}}}

\newcommand{\Msun}{\ensuremath{\mathrm{M}_\odot }}

\newcommand{\twoFone}[1]{\ensuremath{{}_2F_1}}
\renewcommand{\vec}[1]{\ensuremath{\boldsymbol{#1}}}
\definecolor{grey}{rgb}{0.4,0.6,0.6}
\definecolor{brown}{rgb}{0.65,0.16,0.16}
\definecolor{orange}{rgb}{0.9,0.5,0.0}

\title[Feedback from AGN Jets and Quasars]{AGN Feedback Compared: Jets versus Radiation}
\author[S. Cielo, R. Bieri  et al. ]{
\parbox[t]{\textwidth}{
Salvatore Cielo$^{1}$\thanks{E-mail: cielo@iap.fr},
Rebekka Bieri$^{2}$,
Marta Volonteri$^{1}$, 
Alexander Wagner$^{1,\,3}$ and
Yohan Dubois$^{1}$}
\vspace*{6pt} \\
$^{1}$ Institut d'Astrophysique de Paris (UMR 7095: CNRS \& UPMC -- Sorbonne
Universit\'es), 98 bis bd Arago, F-75014 Paris, France\\
$^2$ Max-Planck-Institute for Astrophysics, Karl-Schwartzschild-Strasse 1, Garching \\
$^3$ University of Tsukuba, Center for Computational Sciences (CCS) Ibaraki, Tsukuba, Tennodai 1-1-1
Japan, 305-8577}
\date{Accepted . Received ; in original form }

\begin{document}
\maketitle

\begin{abstract}

Feedback by Active Galactic Nuclei is often divided into \emph{quasar} and \emph{radio} mode, powered by radiation or radio jets, respectively. Both are fundamental in galaxy evolution, especially in late-type galaxies, as shown by cosmological simulations and observations of jet-ISM interactions in these systems.  We compare AGN feedback by radiation and by collimated jets through a suite of simulations, in which a central AGN interacts with a clumpy, fractal galactic disc. We test AGN of $10^{43}$ and $10^{46}$~erg/s, considering jets perpendicular or parallel to the disc. Mechanical jets drive the more powerful outflows, exhibiting stronger mass and momentum coupling with the dense gas, while radiation heats and rarifies the gas more. Radiation and perpendicular jets evolve to be quite similar in outflow properties and effect on the cold ISM, while inclined jets interact more efficiently with all the disc gas, removing the densest $20\%$ in $20$~Myr, and thereby reducing the amount of cold gas available for star formation. All simulations show small-scale inflows of $0.01-0.1$~M$_\odot$/yr, which can easily reach down to the Bondi radius of the central supermassive black hole (especially for radiation and perpendicular jets), implying that AGN modulate their own duty cycle in a feedback/feeding cycle.
\end{abstract}


\section{Introduction}\label{sec:intro}

Supermassive black holes (SMBHs), when accreting matter, become active and produce energy. This energy, in its various forms, couples to the surrounding gas and alters its properties, an effect referred to as ``feedback", from Active Galactic Nuclei (AGN). AGN feedback can thus affect SMBH growth and the global properties of the galaxy over cosmic time.


AGN feedback in the literature is often divided into two ``flavours", a so-called radio-mode powered by mechanical jets, and a so-called quasar-mode powered by photons that couple
to the gas and transfer their momentum. In many situations one dominates over the other. For instance, in radio galaxies the main energy input is kinetic, through jets, while the luminosity is low.  For instance, in M87 the jet power is $\sim 10^{44}$ erg~s$^{-1}$, the X-ray luminosity of the nucleus is  $<10^{41}$ erg~s$^{-1}$. In luminous quasars, the opposite is normally true: high optical/UV/X-ray luminosities are not accompanied by powerful jets. There are occasionally sources with both high radiative power and strong jets, e.g., 3C~273, with  jet power $\sim5\times 10^{44}$ erg~s$^{-1}$, and radiative power $4\times 10^{45}$ erg~s$^{-1}$ \citep{kaspi_reverberation_2000,paltani_bhmass_2005,kataoka_variability_2003}.

Accretion disc and feedback models, including recent numerical work about the origin of jets \citep{sadowski_energy_2013,tchekhovskoy_launching_2015}  predict the two modes to be active under different conditions close to the SMBH, in an ``accretion paradigm"  \citep[e.g.,][]{ulvestad_radio_2001, merloni_fundamentalplane_2003, kording_accetion_2006,merloni_synthesis_2008}.  According to this paradigm, steady jet production occurs when the accretion rate, in Eddington units, is low, $\lesssim10^{-3}$, and the accretion flow is optically thin \citep[geometrically thick,][]{narayan_underfed_1995,abramowicz_advection_1995,blandford_lowaccretion_1999}.  In this state the kinetic power of jets significantly exceeds the radiative power, because the radiative efficiency is typically much lower than the standard  binding energy per unit mass of a particle in the last stable circular orbit.
The opposite is instead expected for thin-disk accretion \citep{shakura_binarybh_1973}, characterising higher accretion rates. In this second state, jet production is suppressed.  


Both radiative and jet feedback  play an important role in quenching galactic SF \citep{tortora_agn_2009,merloni_synthesis_2008} and in powering massive, fast nuclear outflows \citep{tombesi_ufos_2015,feruglio_multiphase_2015} which are part of a galaxy-wide outflow complex, likely powered by the AGN alone.
Large-scale and cosmological simulations of galaxy formation need physically motivated models for either  mode, since including them both (as in the HORIZON-AGN simulation, as described in \citealp{dubois_horizonagn_2016}) has proven successful in reproducing the morphologies of galaxies via regulation with AGN feedback, at the same time recovering observed galaxy luminosity functions and relations between black hole mass and galaxy properties.


However, it is necessary to investigate AGN feedback with high-resolution simulations of idealized galaxies to understand the complex interactions between AGN outflows and a clumpy multiphase ISM and BH feeding. In a recent work, \citep[][henceforth B+17]{bieri_outflows_2017}, studied radiative feedback in isolated gas-rich galaxies. They found that radiative feedback in the energy-conserving regime is capable of driving nuclear and large-scale outflows, highlighting the importance of multi-scattering of IR photons for efficient momentum coupling between radiation and gas.

Feedback by AGN jets in galaxies has been studied in detailed simulations, both in discs (as by \citealp{gaibler_jet-induced_2012}) and elliptical galaxies (e.g. \citealp{gaspari_elliptical_2012,cielo_3d_2014}). Jet/ISM coupling in the latter case is usually efficient because the jets can efficiently percolate through the cold ISM structure and affect the gas in the entire galaxy \citep{wagner_driving_2012}. Sometimes, low-power jets (around $10^{43}$~erg/s), because of their longer confinement time, couple more efficiently than higher power jets \citep[e.g., ][]{mukherjee_relativisticdynamics_2016}.

Observations of jets interacting with individual ISM clumps are well documented in the literature. Interactions occur even in the central few hundreds of parsecs of spiral galaxies such as NGC4151 (\citealp{wang_jetcloudcollision_2011}), or NGC4258 (\citealp{wilson_NGC4258_2001} and \citealp{cecil_jets4258_2014}), although jet feedback in disc galaxies is reputed less efficient due to the high jet directionality. In late-type galaxies misaligned jets can sometimes lie in or very close to the disc plane (\citealp{battye_radioorientation_2009} and \citealp{browne_jetorinentation_2010}, as discussed also by \citealp{dubois_bhevolutionIII_2014}).
Misalignments between the disc plane and the inner AGN structure are also observed, as in the Megamaser disc measurements by \citet{greene_megamaser_2013}.
These findings show that jet systems can have a large  impact on galactic discs at all scales. 

Radiative and jet feedback are expected to couple differently with the interstellar medium (ISM), but the differences have not been systematically quantified yet. In this work we compare radiative and jet feedback in a galactic disc, at fixed power, lifetime and initial conditions. For jet feedback, we investigate the cases of jets both perpendicular and parallel to the disc plane. 

In our AGN models we measure:
\begin{itemize}
\item the mechanical coupling to the ISM and the outflow rates;
\item the effects of the AGN on the central region, including small-scale inflows and consequences on duty cycles;
\item the global effect on the gaseous disc, with implications for the SFR in high-z galaxies.
\end{itemize}

In Section \ref{sec:setup} we describe the physics and the implementation of our simulation of our simulation setup, while in Section \ref{sec:evolution} we present our simulation runs, and give an overview of the evolution of each AGN, comparing  visually the outflows and the effects on the disc gas. In Section \ref{sec:outflow} we quantify the outflow rates and physical state, as well as the mechanical coupling between the AGN and the dense phase of our ISM. 
All our runs exhibit significant inflows towards small scales (hundreds down to tens of parsecs) which we characterize in Section \ref{sec:inflows}. 
Finally, in Section \ref{sec:fb} we investigate the effects of feedback on the cold gas phase, both by zooming on a single representative clump and on the global density distribution function.


\section{Numerical methods}\label{sec:setup}
We are interested in comparing the effects of a radiatively-driven AGN wind and of a jet plasma on the state and dynamics of the cold phase material in the galactic disc. We thus perform a suite of simulations using the hydrodynamical adaptive mesh refinement (AMR) code \Ramses\ \citep{teyssier_ramses_2002}. 
\Ramses\ solves the classical Euler fluid equations coupled with an equation of state for ideal monoatomic gas with an adiabatic index of $\gamma = 5/3.$ We keep the simulations' physics as simple as possible in order to isolate the effects of radiation or the jet. We do not include gravity (neither a static gravitational field nor self-gravity), or radiative cooling losses from the gas. No additional (subgrid) astrophysical process such as star formation or supernovae feedback are included either.


\Ramses\ \citep{teyssier_ramses_2002} follows the evolution of the gas with a second-order unsplit Godunov scheme. We use the HLLC Riemann solver \citep{toro_HLLriemann_1994} with MinMod total variation diminishing scheme to reconstruct the interpolated variables from their cell-centered values. 

To model the interaction of radiation from the central black hole with the galaxy's gas we run our simulations using \RamsesRT\ \citep{rosdahl_ramsesRT_2013,rosdahl_M1_2015}, a radiation-hydrodynamic (RHD) extension of \Ramses. 
\RamsesRT\ computes the propagation of photons and their interaction with hydrogen and helium via photoionisation, heating, and momentum transfer, as well as their interaction with dust particles via the transfer of momentum self-consistently and on-the-fly. The advection of photons between the grid cells is described with a first order moment method, where the set of equations is closed with the M1 relation for the Eddington tensor.  The method used accounts for the diffusion of multi-scattering IR radiation and is described in detail in~\citet{rosdahl_M1_2015}. Since the \emph{Courant-Friedrichs-Lewy} (CFL) condition imposes a time-step  (and thus the computational load) that scales inversely with the speed of light $c$, we adopt a reduced speed of light approximation $c_\mathrm{red} = 0.2c$ (see also \citealp{gnedin_eddington_2001} and \citealp{rosdahl_ramsesRT_2013} for a more detailed explanation and \citealp{bieri_outflows_2017} for a discussion on the used reduced speed of light).

\subsection{Mesh, resolution, refinement}
\label{subsec:Ramses}

As in B+17 we implement outflow boundary conditions, where any matter that leaves the simulation volume is assumed to be lost to the system. By choosing a sufficiently large box size of 96~kpc we ensure that the mass loss is negligible for most of the simulation. The simulations have been performed with a spatial resolution of $\Delta x = 5.8$~pc for the minimum cell size, corresponding to a maximum refinement level of~14, for all of the simulations. The coarsest level of the simulation is set to 9, corresponding to a spatial resolution of 187.5~pc. The refinement is triggered with a quasi-Lagrangian criterion ensuring that if the gas mass within a given cell is larger than $10^{4}$~M$_{\odot}$ a new refinement level is triggered. The refinement criterion is chosen such that at the beginning of the simulation the whole disc is maximally refined, whereas the hot circum-galactic medium is resolved with a spatial resolution of $\Delta x = 11.6 - 23.2$~pc. 

For the jet simulations we add also two additional custom refinement criteria, based on pressure and total energy (see Section \ref{ssub:jets}).

\subsection{Initial gas density distribution}
\label{subsec:DensityDist}

We initialize the gaseous disc of our simulation with a two-phase ISM in initial pressure equilibrium, following the approach detailed in \citet{sutherland_interactions_2007}. The disc is the same one used by B+17 to investigate the interaction of a radiatively-driven AGN wind with the non-uniform ISM and its dynamical effects. A similar ISM model, but with a spherical distribution of dense clouds, has also been used by \citet{wagner_relativistic_2011} to study the effect of an AGN jet on the ISM. 

We first set up an isothermal uniform hot phase with temperature $T \sim 5\times10^6$~K, in which we embed a cold $T \sim10^4$~K turbulent, inhomogeneous gas distribution.  As in B+17, we set up the disc to mimic a typical compact, gas-rich, high-redshift galaxy, similar to what is found in observations (e.g. \citealt{daddi_very_2010}). 

The density field of the cold phase is uniform on large scales, but very clumpy on intermediate to small scales. The gas obeys a single-point log-normal density distribution and two-point fractal statistics. The two-point structure of a homogeneous turbulent medium is characterised in Fourier space by an isotropic power spectrum $D(k)$ defined as
\begin{equation} 
D(k) = \int 4 \pi  k^2 F (\vec{k}) F^* (\vec{k}) dk \propto k^{-5/3} \quad ,
\label{eq:fractal}
\end{equation}
where $F (\vec{k})$ is the Fourier transform of the density field $\rho (\vec{r})$. 

The fractal density distribution is created using the algorithm described by \cite{lewis_jp4.16_2002} and implemented in \textsc{pyFC}\footnote{https://pypi.python.org/pypi/pyFC}. It generates the 3-dimensional random density distribution that simultaneously satisfies single-point log-normal statistics and two-point fractal statistics. The statistical properties are consistent with those observed in molecular clouds in our Galaxy \citep{kainulainen_probing_2009,roman-duval_mc_2010}; as well as simulations of self-regulated multi-phase ISM turbulence \citep{wada_norman_multiphaseISM_2001}. 

Following \citet{bicknell_jet-induced_2000} and \citet{wagner_relativistic_2011} we adopt a mean value of the log-normal density field of
$\nu = 1$ and a variance of $\sigma ^2 = 5$. In Fourier space the density field follows the Kolmogorov power-law self-similar structure with index $-5/3$ shown in Equation \ref{eq:fractal} with a minimum sampling wave-number $k_\mathrm{min}$. In real space, the minimum sampling wave-number determines the scale of the largest fractal structure in the cube relative to the size of the cube and is chosen to be 5~kpc$^{-1}$ in the simulations presented here (corresponding to the \textit{medC} simulations in B+17). The chosen values are in agreement with ranges found by \citet{fischera_starburst_2003} and \citet{fischera_column_2004}.

It is important to stress that the setup is stationary and, thus, does not capture the actual ISM turbulence dynamically (e.g., \citealp{kritsuk_density_2011}, and references therein). Instead, the initial conditions represent one instantaneous realisation of the non-uniform properties of a generic turbulent medium, relying on a range of previous experimental and theoretical results from the field of turbulence. The adopted initial conditions, characterized by the variance of the gas density and the two-point self-similar power-law structure, can therefore be regarded as a physically motivated generalisation of an inhomogeneous ISM, as explained in \citet{sutherland_interactions_2007}.

Following B+17, the cube is then placed into the \Ramses\ simulation domain by reading each density value within the cube and placing it into the \Ramses\ grid. The density cube is further filtered, in the $xy$-plane, by a symmetric flat mean density profile with mean cold phase density $\langle n_\mathrm{w} \rangle$ and radius $r = 1.5$~kpc. In the $z$-plane we filtered the density cube by a step function with height $h=0.3$~kpc. 
We ensured that the cube is placed into the simulation domain in such a way that the cloud with the highest central density is centered in the galaxy, thus placing the AGN into the densest environment of the entire density distribution.

\begin{figure}
 \includegraphics[width=\columnwidth]{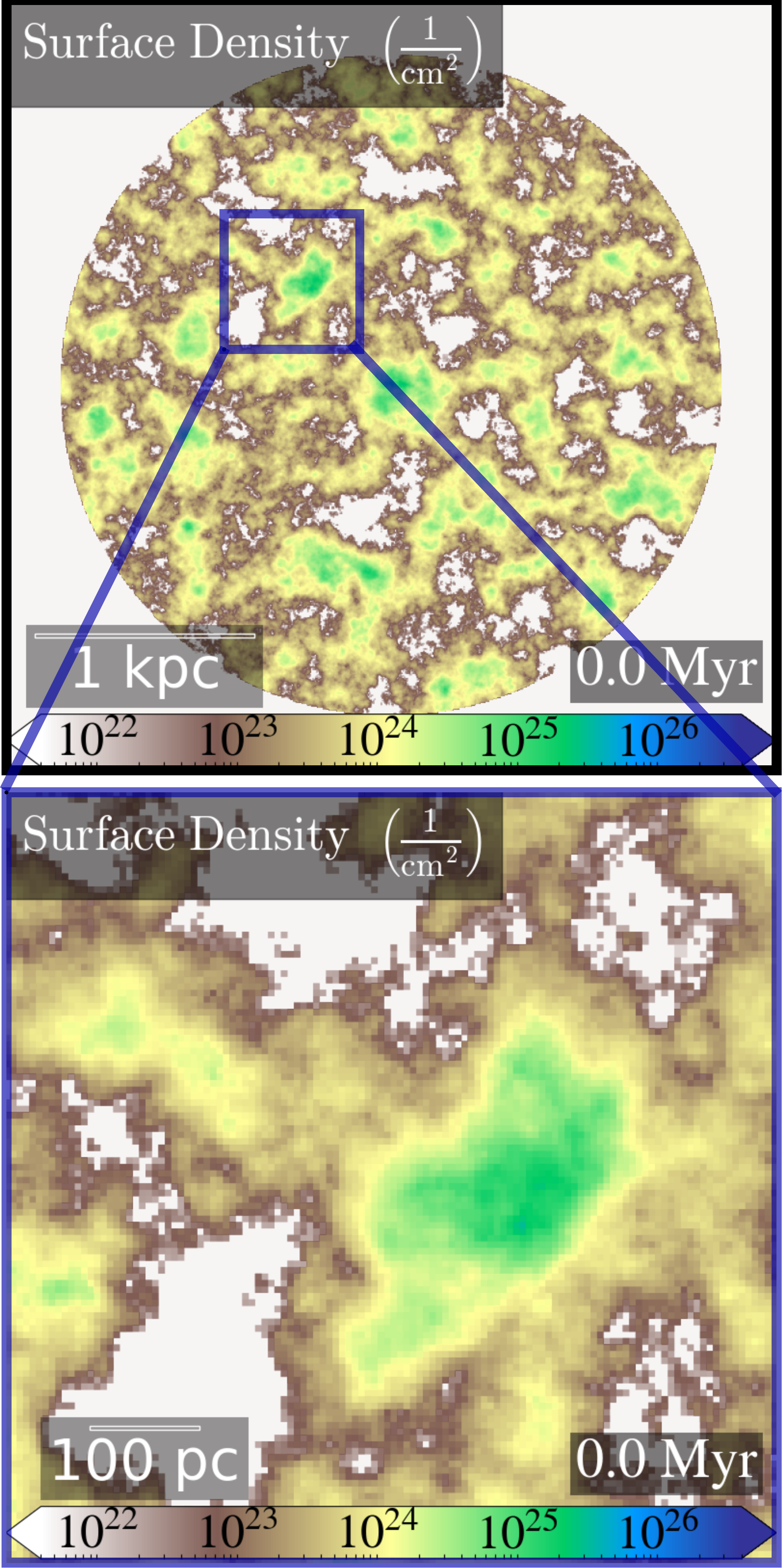}
  \caption{Common initial conditions for all our runs. Face-on image of the gas surface density (in $cm^{-2}$) for the full simulated disc ($3.5$~kpc $\times3.5$~kpc view) and our chosen zoom-in clump ($0.5$~kpc $\times0.5$~kpc view). }
  \label{fig:IC}
\end{figure}

Furthermore, the porosity of the ISM arises by imposing a temperature roof for the existence of the cloud $T_\mathrm{roof}$. The mean density (around 500~H~cm$^{-3}$) of the cold ISM phase and the roof temperature (70~K) is chosen such that the total cold gas mass within the galaxy is $\sim 2 \times 10^{10}$~\Msun.
Figure \ref{fig:IC} shows the face-on surface density for our initial conditions.

The cloud cells are initialized with solar metallicity. The metallicity is modeled in \Ramses\ as a passive scalar for the gas advected with the flow. In our simulations, the metals are a fundamental ingredient in our dust model (see section \ref{sub:qso}, but we are able to use them as well for post-processing estimates of the gas' cooling time (section \ref{sec:fb}).
The clouds are assumed to be initially in pressure equilibrium with the surrounding hot phase, with the pressure set to be $p_\mathrm{h} \sim 7 \times 10^{-12}$~Pa. The density and temperature of the hot phase is constant and set as $\rho _\mathrm{h} = 0.01$~H~cm$^{-3}$ and $T_\mathrm{h} \sim 5\times10^{6}$~K, respectively, for all the simulations. The filling factor of the cold phase within the disc is given by 
\begin{equation}
f_\mathrm{V} = \frac{V_\mathrm{cold}}{V_\mathrm{tot}} \quad ,
\end{equation}
where $V_\mathrm{cold}$ is the volume of the cold phase and $V_\mathrm{tot}$ is the total cylindrical volume of the cylindrical volume within which the density is distributed in. For the simulations presented here we have chosen a filling factor of 50\%, as what has been used for the majority of the simulations in B+17.

\subsection{Timescales}
Note that in the simulations presented we neglect gravity and gas cooling.  The two important timescales associated with these two effects are the cloud collapse time, equivalent to the cloud free-fall time, and the cooling time in different environments. 

\begin{table*}
\begin{center}
\caption{The photon groups used in the simulations are defined by an energy
interval $\Delta E$. The adapted cross sections, taken from
\citet{verner_crosssections_1996}, for ionisation of hydrogen and helium are shown in the
next three columns. Each photon group has a given dust opacity $\tilde{\kappa}$. The designated energy from the AGN that goes into
each corresponding photon group is calculated using the quasar spectrum adopted
in \citet{sazonov_spectrum_2004}. }
\begin{tabular}{llcllllrc}
\specialrule{.12em}{.05em}{.05em}
Photon group & \multicolumn{3}{c}{$\Delta E$ [keV]} &
$\sigma_{\mathrm{H}_{\mathrm{I}}}$ [cm$^2$] &
$\sigma_{\mathrm{He}_{\mathrm{I}}}$ [cm$^2$] &
$\sigma_{\mathrm{He}_{\mathrm{II}}}$ [cm$^2$] & $\tilde{\kappa}$
[$\mathrm{cm}^2\,\mathrm{g}^{-1}$] & Energy fraction per group\\
\specialrule{.12em}{.05em}{.05em}
IR      & [1e-5    &-& 0.001]     & 0                   & 0                   & 0                   & 10    & 0.310   \\
Opt     & [0.001   &-& 0.0135]    & 0                   & 0                   & 0                   & 1000  & 0.250   \\
UV1     & [0.0135  &-& 0.02459]   & $3.1\times10^{-18}$ & 0                   & 0                   & 1000  & 0.079   \\
UV2     & [0.02459 &-& 0.0544]    & $4.7\times10^{-19}$ & $4.2\times10^{-18}$ & 0                   & 1000  & 0.067   \\
UV3     & [0.0544  &-& 1.]        & $1.1\times10^{-20}$ & $2.3\times10^{-19}$ & $1.7\times10^{-19}$ & 1000  & 0.085   \\
\specialrule{.12em}{.05em}{.05em}
\end{tabular}
\label{tab:rtinit}
\end{center}
\end{table*}

The cloud collapse time is defined as 
\begin{equation}
t_\mathrm{ff} = \frac{1}{\sqrt{G \rho_\mathrm{c}}}
\end{equation}
where $G$ is the gravitational constant and $\rho_\mathrm{c}$ is taken to be the typical cloud density. With a value of $10^2 - 10^3$~H~cm$^{-3}$ the cloud collapsing time is 3-9~Myr, justifying neglecting the gas' self-gravity, over large part of our simulations. The orbital time of the disc is also longer than the simulation time.


The cooling time for a gas with temperature $T$ and particle number density $n$ is defined as
\begin{equation}
t_\mathrm{cool} = \frac{k_B T}{n \Lambda(T)}
\end{equation}
where $\Lambda(T)$ is the cooling function employed in \Ramses. Various cooling models are implemented into \Ramses\ but the most common cooling model uses the standard H and He cooling by Katz et al. (1996), with an additional contribution from metals, dominant above $10^4$~K \citep{sutherland_cooling_1993}, and molecular gas at lower temperature (e. g. \citealp{vasiliev_cooling_2013}). 

With a temperature of $5\times10^6$~K the cooling time of the hot ISM exceeds the Gyr and is thus longer than the simulation time. The cooling time in the clouds, with a temperature of $\sim 10^1 - 10^2$~K and densities between $10^2 - 10^3$~H~cm$^{-3}$ is instead between $\sim$~3-40~kyr and thus short enough to affect the temperature evolution of the cloud. The presence of feedback can  significantly increase the cooling time of the dense gas, as we show in Section \ref{sec:fb}.

The addition of gravity and cooling in the presented simulations would lead to two effects. First, cooling enhances the life-time of clouds \citep{cooper_massmetallicity_2008}. This alone decreases the feedback efficiency, as diffuse gas is easier to accelerate than condensed cold gas. Second, the inclusion of cooling possibly causes the fragmentation of the massive clouds leading to a building of lower density channels around the fragments that would further weaken the mass loading  efficiency of the AGN (see \citealp{bieri_outflows_2017} for a detailed explanation). 
Gravity and cooling would cause formation of dense cores with a larger optical depth, to which the radiation could couple more efficiently. 

The inclusion of gravity and cooling increases the complexity of the non-linear interplay between the AGN feedback and the gas, and we defer such study to future work.

\subsection{AGN Models}\label{subsec:agnparams}

\subsubsection{Modeling of the radiation from a quasar}\label{sub:qso}
We model the spectrum of the radiation from the black hole by five photon groups, defined by the photon energy intervals $\Delta E$ listed in Table~\ref{tab:rtinit} and taken identical to B17+. For each individual photon group the radiative transfer equations are solved separately. The spectrum is divided into one infrared (IR) group, one optical group, and three groups for the ionising ultraviolet (UV) photons.

From the broad-band spectral energy distribution adopted in \citet{sazonov_spectrum_2004} we calculate the corresponding fraction of the energy distribution, which is then multiplied by the quasar quasar luminosity to yield the corresponding group energy for each photon group.  In these simulations we do not model the X-ray energy band, whose photons are thought to have  little impact on the kpc-scale ISM \citep{ciotti_elliptical_2012, hopkins_concert_2016}, except very close to the central SMBH.  The interaction of the UV photons with hydrogen and helium via photoionisation is determined by the ionisation cross sections $\sigma (E)$ taken from \citet{verner_crosssections_1996}.

In Table~\ref{tab:rtinit} the photon energy intervals $\Delta E$, photoionisation cross-sections ($\sigma_\mathrm{HI}$, $\sigma _\mathrm{HeI}$, and $\sigma_\mathrm{HeII}$) for ionisation of hydrogen and helium, the dust-interaction opacity ($\tilde{\kappa}$), and energy fractions used per photon group is listed.

The interaction of the IR and optical photons with the gas happens via the dust and is determined by the dust opacity $\tilde{\kappa}$ that scales with the gas metallicity as $\tilde{\kappa} _{\rm IR} Z / Z_\odot = 10
\, Z / Z_\odot \rm{cm}^2 \rm{g}$. For the IR radiation the dust opacity is
additionally multiplied by $\exp(-T/T_{\rm cut})$ to account for thermal dust
destruction by sputtering, with $T_{\rm cut}=10^5$. This is particularly necessary when the source luminosity is sufficiently high to heat the gas to very high temperatures above which it would be unrealistic for dust to still exist. For a more detailed discussion see B+17. Note, that by initializing the cold phase of the ISM with solar metallicity we may overestimate the effect of dust for a very high-redshift galaxy. 


\subsubsection{Modeling of the jets}\label{ssub:jets}
The jets are introduced as classical hydrodynamic source terms, through essentially the same setup as described in \citet{cielo_backflows_2017} (henceforth C+17), but now implemented in \Ramses\ and with adjustable orientation. First, the jet direction is parametrized by two Euler-angle parameters, $\alpha_\mathrm{jet}$ (inclination with respect to z-axis), and $\beta_\mathrm{jet}$ (angle in the meridional plane). The circular faces of a cylinder in the chosen direction, covering a total area $A_\mathrm{jet}$ are chosen as jet bases. A two-cell gap is left between the opposite jet bases to ensure convergence in \Ramses\ Riemann's solver.

A jet density $\rho_\mathrm{jet} = 0.01 \rho_\mathrm{h}$ is chosen, in order to have a light, mildly supersonic jet. The corresponding mass and momentum flux are then calculated and reported in Table \ref{tab:sim}, together with the values for the radiative runs.


The high outflow speed of this classical hydro source term poses some strict limits the time-step calculation trough the aforementioned CFL condition.  The condition gets computationally very demanding along the jet beam, when it is maximally refined, as at high power the jets' velocity approaches a considerable fraction of the speed of light. 
Actual jet beam velocities are often slowed down by entraining surrounding gas, so typical velocities are of order
$60000$~km/s. This is an advantage, but after the first interactions, velocities at the jet base 
remain still very high, but lowering the refinement level in the hot phase can alleviate the demand.

This is one of the reasons why in the jet simulations we make use of additional refinement criteria, based on fixed fractions of the jet pressure and total energy at injection. The main advantage of this approach is that the jet pressure and total energy are always defined in the code, so no additional variable has to be defined (introducing additional memory requirement) or calculated at every timestep. 

In detail, the pressure and total energy at the base of the jet ($p_{base}$ and $e_{base}$) are recorded during the whole jet lifetime. When the jets are switched off (see below for the lifetime of our AGN), the last recorded values are assumed. At any point in the simulation, any computational cell is refined whenever its pressure or total energy exceed a value of $f_\mathrm{p}\ p_{base}$ or $f_\mathrm{e} \ e_{base}$, respectively. 

The values for $f_\mathrm{e}$ and $f_\mathrm{p}$ are chosen as follows. 
At the beginning of the simulations we set both to $0.05$; this value allows very reliably for full refinement of the jet beam (through $f_\mathrm{e}$) and terminal shocks and the resulting hot bubbles (through $f_\mathrm{p}$). Later in the simulation (about $3.5$~Myr in p43 and $1.5$ in p46), when a considerable fraction of the shocked gas leaves the disc plane, we raised both values to $10$, in order to lower otherwise prohibitive computational costs caused by high refinement levels in regions of hot, jet-entrained outflows immediately outside the beam or terminal shocks. Instead, these regions were refined at most to level 12 if far from the disc plane; for our simulations, high resolution in the hot phase is not required. After the energy sources are switched off, the custom criteria continue to apply, but no high refinement level is reached, and the result is insensitive to the values of $f_\mathrm{e}$ and $f_\mathrm{p}$ used. Thus, the parameters were then set to $10000$ to avoid possible refinement triggering in the hot outflows very far from the disc. For jets starting after the beginning of the simulation, a condition for a CFL time-step limiter taking place a few timesteps before the jets has to be added.

We present three distinct jet models per power, differing only by jet orientation: in one model, the jets are perpendicular to the disc plane ({ver}), while in the other models, the jets lie in the plane of the disc and are directed along the directions of maximum/minimum inertial mass ({max} and {min}). We calculated the direction of maximum/minimum inertia by integrating the cell mass along 500 randomly selected rays in cylindrical radius along the disk plane, uniformly sampling along a ray from the centre of the disc, where the AGN is located, up to a distance of 1.5~kpc. The jet directions are listed in Table \ref{tab:sim} next to all the other parameters used for the simulation suite.

%

\subsection{Simulation parameters summary}
Table \ref{tab:sim} summarizes the parameters of our simulations.  The 8 simulations are divided into two groups that differ in AGN power and lifetime. Each of the two groups is further divided by the type of energy injection (radiative or mechanical) and jet inclination with respect to the disc (vertical or along the path of minimum/maximum column density in the disc). 


For an unbiased comparison, we adopt the same power and lifetime for both the radiative, a.k.a. quasar, and jet models. After B+17, we adopt two powers, $10^{43}$ and $10^{46}$~erg/s (p43 and p46), representing faint and bright AGN. We then calculate appropriate lifetimes $t_\mathrm{on}$ from observational literature, choosing $t_{on}$ as $10$~Myr (as in B+17) and $1.67$~Myr for low and high power, respectively. These timescales  are derived by fitting the power/age relation for radio sources from \citet{parma_age_1999} (the same relation adopted in C+17), and are thus mostly appropriate for jets. 
The chosen ages are also consistent with the range of values found by \citet{turner_lifetimes_2015} for their volume-limited sample of local radio galaxies.
All simulations are evolved up to 20 Myr.

\begin{center}
\begin{table*}
\caption{Parameters of different simulations: name, AGN power ($P_{AGN}$), mass loading and input momentum flux, AGN model (radiative or mechanical), jet inclination angles (polar, azimuthal; for jets only), AGN lifetime $t_{on}$.}
\tabcolsep 3pt
\begin{tabular}{lcccccc}
\hline
Run name  & $\log P_{AGN}$            & Input mass loading       &  Input momentum flux     &  AGN  Model & Inclination    & $t_{on} $    \\
          & (erg$\,$s$^{-1}$)   & (M$_\odot$/yr) & (M$_\odot$/yr km/s)           & $(\alpha, \beta)\deg$     & Myr        \\
\hline
{\bf p43 series} & & & & & & \\
\hline
\verl     & 43 & $3.5\times10^{-4}$ & $73.3$ & jet perpedicular to disc & (0, 0)    & 10  \\
\minl     & 43 & $3.5\times10^{-4}$ & $73.3$ & jet in disc, min inertia & (90, 44)  & 10  \\
\maxl     & 43 & $3.5\times10^{-4}$ & $73.3$ & jet in disc, max inertia & (90, 140) & 10  \\
\qsol     & 43 & $0$                & $52.9$ & radiative                & N/A       & 10  \\
\hline
{\bf p46 series} & & & & & & \\
\hline
\verh     & 46 & $7.5\times10^{-2}$ & $3.4\times10^4$ & jet perpedicular to disc & (0, 0)    & 1.67  \\
\minh     & 46 & $7.5\times10^{-2}$ & $3.4\times10^4$ & jet in disc, min inertia & (90, 44)  & 1.67  \\
\maxh     & 46 & $7.5\times10^{-2}$ & $3.4\times10^4$ & jet in disc, max inertia & (90, 140) & 1.67  \\
\qsoh     & 46 & $0$                & $5.3\times10^4$ & radiative                & N/A       & 1.67  \\
\hline
\label{tab:sim}
\end{tabular}
\end{table*}
\end{center}

\section{Simulation Evolution and Appearance} \label{sec:evolution}

\begin{figure*}
  \includegraphics[width=\textwidth]{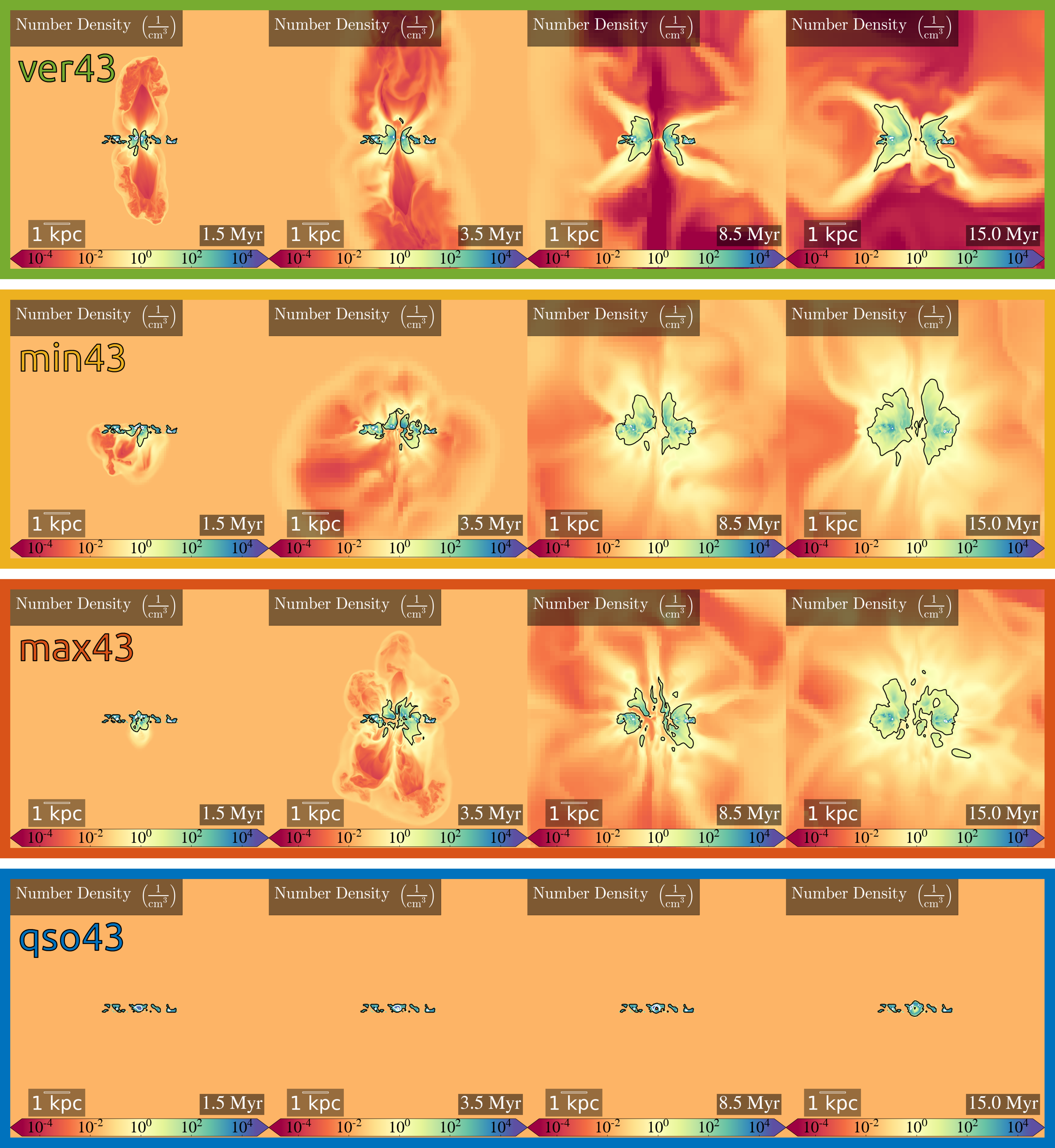}
\caption{Number density evolution for all the p43 runs (from top to bottom: vert, min, max, qso). Each panel is a $10$~kpc-sided slice along the x = 0 plane; showing $1.5$, $3.5$, $8.5$ and $15$~Myr (from left to right). The superimposed contours trace isodensity surfaces at $4$ (black) and $2500$~cm$^{-3}$ (white).
}\label{fig:dens43}
\end{figure*}
\begin{figure*}
  \includegraphics[width=\textwidth]{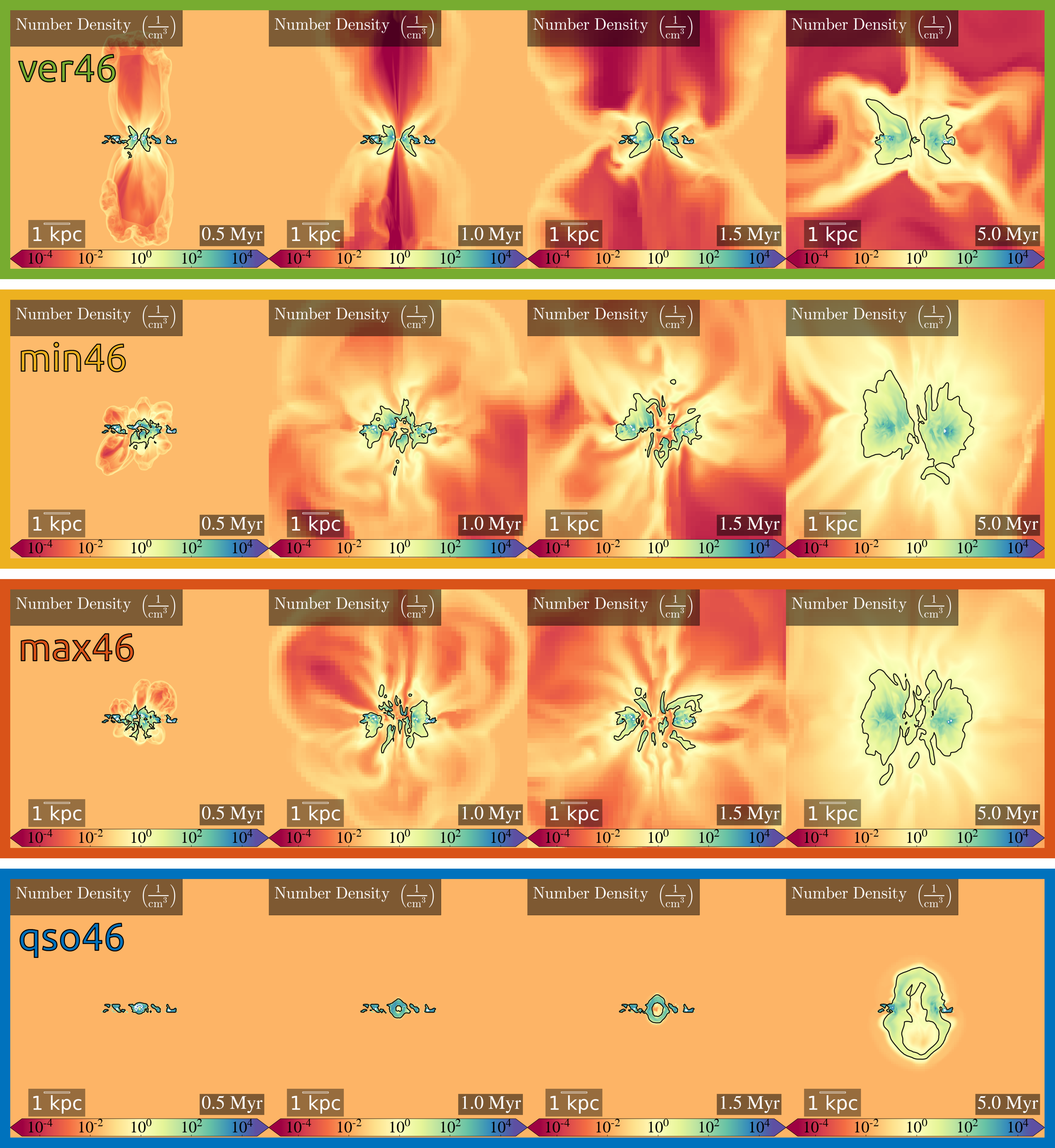}
  \caption{Same as Figure \ref{fig:dens43}, but for the p46 runs, snapshots at $0.5$, $1.0$, $1.5$ and $5$~Myr. Note the similar yet faster evolution for the p46 jets while the jet is still on, and the much more extended outflows afterwards.}\label{fig:dens46}
\end{figure*}
\begin{figure*}
  \includegraphics[width=\textwidth]{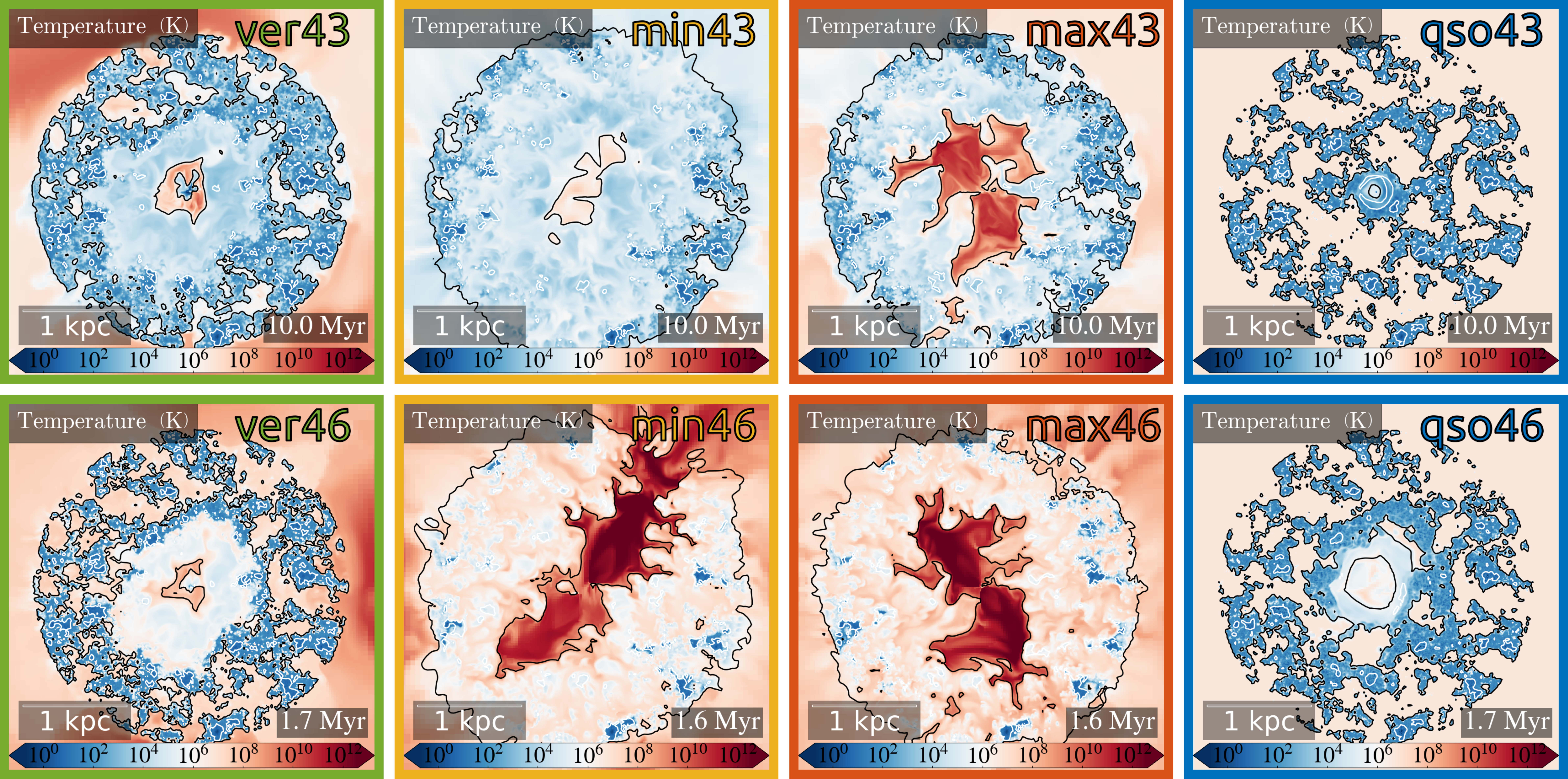}
  \caption{Temperature slices for all runs, along the $z=0$ plane, right before the AGNs are switched off. The scale is $3.5$~kpc. The contours trace isodensity surfaces at $4$ (black) and $2500$~cm$^{-3}$ (white). The jet inclination in the min and max models is approximately along the first and second diagonal, respectively. }\label{fig:tempfo}
\end{figure*}

We begin by describing the appearance of the outflow and the visible effects on the disc for our eight AGN models. Figures \ref{fig:dens43} and \ref{fig:dens46} show number density slices along the $x=0$ plane (edge-on view of the disc), $10$~kpc side, for the p43 and p46 runs, respectively. Each row follows the time evolution of one run. Figure \ref{fig:tempfo} shows a $z=0$ slice (i.e. going through the middle of the disc plane) of the gas temperature, always at the last snapshot in which the AGN is on. By comparing these figures we can see how the outflows interact with the gas in the disc.

After the vertical jets leave the disc, lobe structures of a few kpc length are created for both ver43 and ver46 simulation. The jets entrain large amounts of gas from the ISM around the AGN which is expected, since the jet first has to break out the initial overdensity. As a result, the jet beam is slowed down, the reconfinement shocks are made less efficient making the beam and its terminal shock broader. Similar mechanisms have been described by \citet{perucho_deceleration_2014}. These broadened jet beams end in a Mach disc (i.e. an extended, circular terminal shock as wide as the beam) a fraction of a kpc in diameter, rather than in a point-like hot spot (such as jets with the similar parameters do in C17, in absence of cold gas).

Vertical jets also heat the central kpc of the disc very effectively, resulting in large amounts of warm, expanding gas that is expelled from the disk plane.  The jet material then expands and creates two large, hot bubbles (passing first through a classical radio-lobe phase in the case of ver43). At this stage, the bubbles still interact with the disc mid-plane, but a few Myr after the switch-off, they detach from the disc and get usually de-refined.

In the case of the min and max runs, the outflows they power outside the disc plane look very similar to the case of vertical jets. This is mostly true at high power, since high power jets in the disc plane interact globally with the disc material (directly through the main jet beam, or indirectly trough secondary jet streams and accelerated, shocked gas). 
In all cases, the jet streams must flood through channels in the cold ISM, leading to the inter-cloud channels of the disc being filled with hot shocked gas. The jet beam is deflected away from the disc plane by interacting with a cold clump and the cold ISM is disrupted in several fast-outflowing components. The former behaviour is more common at low power, the latter at high power, though to some degree both can be observed in either series. This effect was also seen in the evolution of an accretion disc wind propagating through a disc \citep{wagner_ultrafast_2013}.

After $1$ or $2$~Myr, the jets have replaced the hot ISM inside the disk with their hotter, higher pressure plasma; for the vertical jets however this does not extend past the central kpc. This overpressure compresses the gas slightly (as modeled for instance by \citealp{bieri_positivefb_2015} and seen in the simulations by \citealp{gaibler_jet-induced_2012}) but it is also effectively displacing the gas, some of it inwards and some of it outwards, along the disk plane (see Section \ref{sec:inflows}).

Although not directly visible in the density plot, the photons in  qso43  manage to push the gas outwards and carve a central hole into the initial clump.  However, due to the high central density, the photons never manage to break up the central clump and the outflow stays confined within it.  Most of the IR radiation  leaks out and escapes the galaxy mildly heating gas at larger distances, but most of the total energy remains in the radiation field, with minimal coupling to the gas.

Comparing the corresponding simulations in the p43 and p46 series, the hot and warm gas phases show  an overall similar evolution, heating the same volumes of gas. The evolution is  faster in p46, and the gas is heated to higher temperatures. The most notable exception is run qso46, whose central region is affected by the AGN much more than in its inefficient p43 counterpart, and evolves (after about $12$~Myr) to be more similar to run ver46 in appearance and properties properties of the outflows, especially towards the end of the simulations, when the radiation-heated gas breaks out  of the dense shell that was confining it.

At first, all jets interact exclusively with the central clump, while in the radiative runs some IR radiation can escape even at early times, heating other clumps at larger radii.
Most of our simulated jets first split the central clump in two through ram pressure and shocks driven into the clump. 
The way the jets interact with the two halves afterwards depends on the jet's direction and power. At low power, the jet in min43 is easily deflected outside the disc plane by collision with the remnants of the central clump. This happens further down the jet path for run max43. At high power, the jet beam is not as easily deflected, but it is instead more often split by the clumps into several fast outflowing streams. 
In both ver43 and ver46, the interaction with the central clumps
determines, besides the mass loading of the jet, asymmetric bending of the beam right at its base, and a long-lasting ``flickering'', so that jets seem to undergo a few bursts whenever gas is entrained, as visible from Figures \ref{fig:dens43} and \ref{fig:dens46}.
Similar jet activities have been observed by \citet{homan_mojavejetbase_2015}, in which self-collimating AGN  jets show sudden sharp velocity changes on $10$ to $20$~pc scales, hints at interaction of the jet stream with dense material at the centers of galaxies.

In paricular, the ver43 case has also its jet beams bent by a few degrees, both towards the positive values of the y-axis.
Thus the interaction of the beam at small radii with a dense ISM alone may be responsible for a similar asymmetric bending in a real jet beam, without necessarily implying a bulk velocity of the BH with respect to the hot ISM, or the interaction of the beam with magnetized stellar winds.

As for the qso runs, qso43 is an example of a radiative outflow confined within the central overdensity for all its lifetime, but also run qso46 has to spend considerable energy to break out of it, as the driving of outflows outside the disc plane is delayed by a few Myr compared to the jetted cases. No flickering is observed in the outflows from the radiation/ISM interaction, but we observe piecemeal, irregular secondary inflows (section \ref{sec:inflows}) which may cause short bursts in the AGN over a few Myr timescales.

\section{Outflows and acceleration}\label{sec:outflow} 

\begin{figure*}
  \includegraphics[width=\columnwidth]{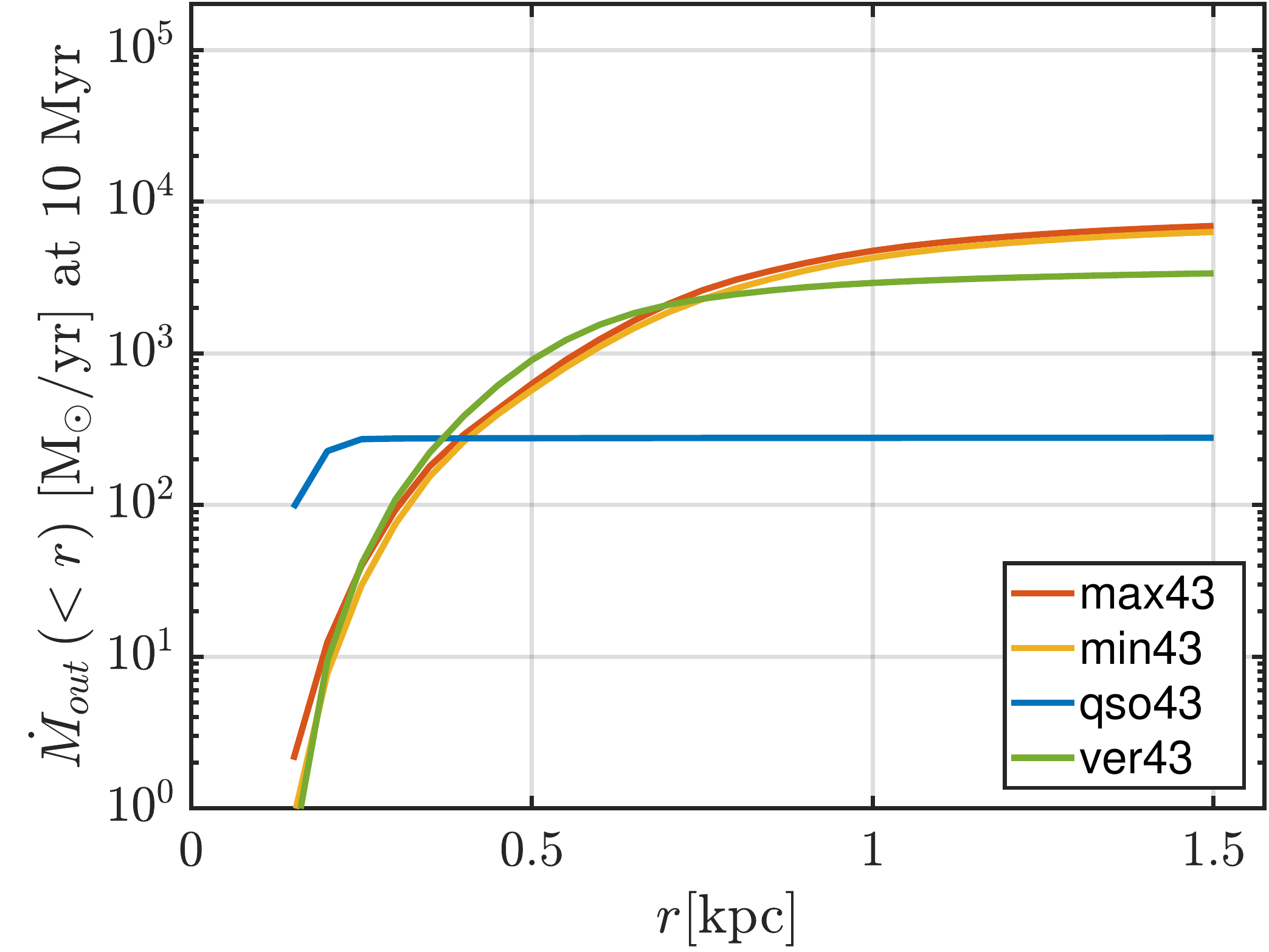}
  \includegraphics[width=\columnwidth]{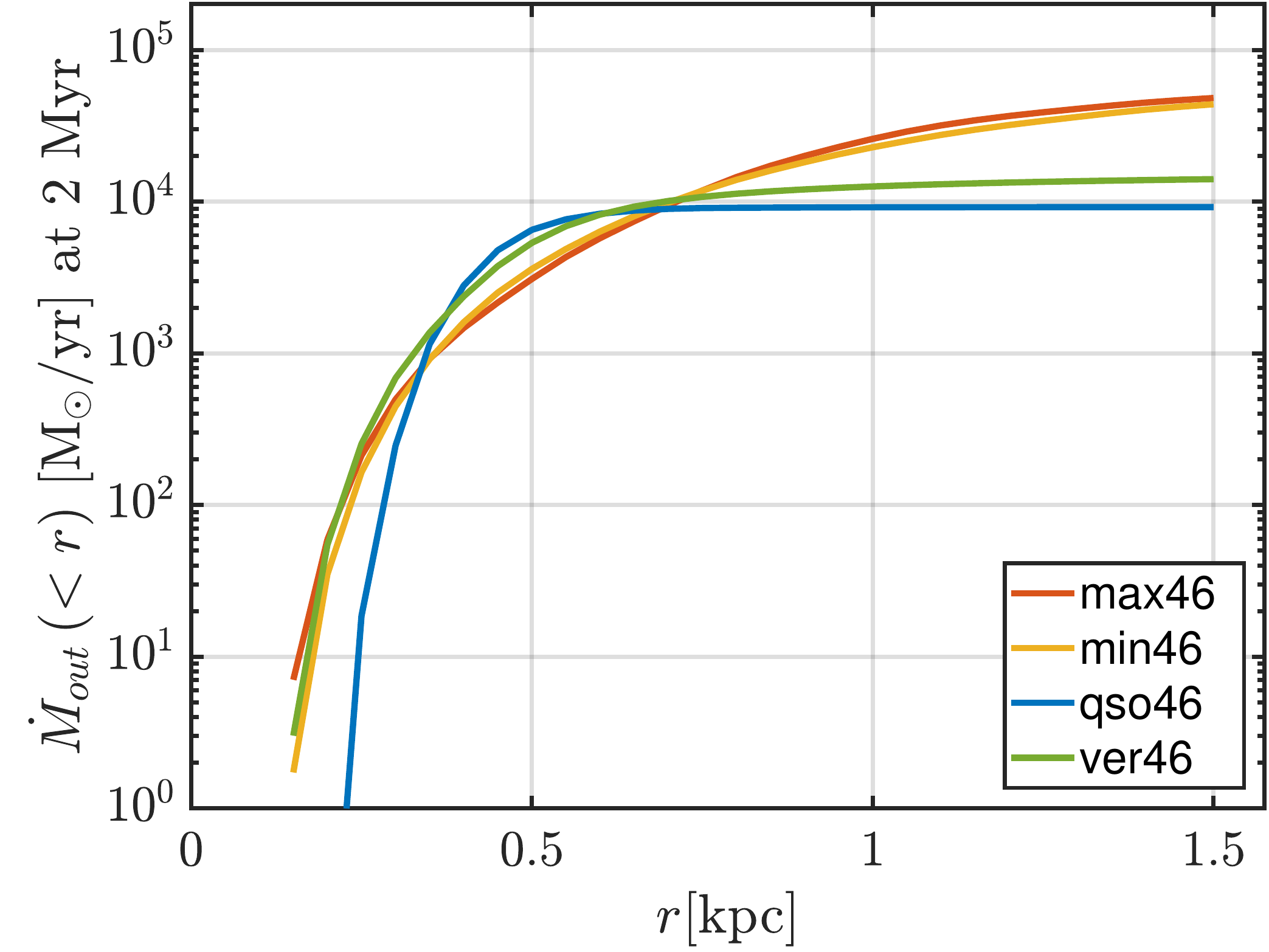}
  \includegraphics[width=\columnwidth]{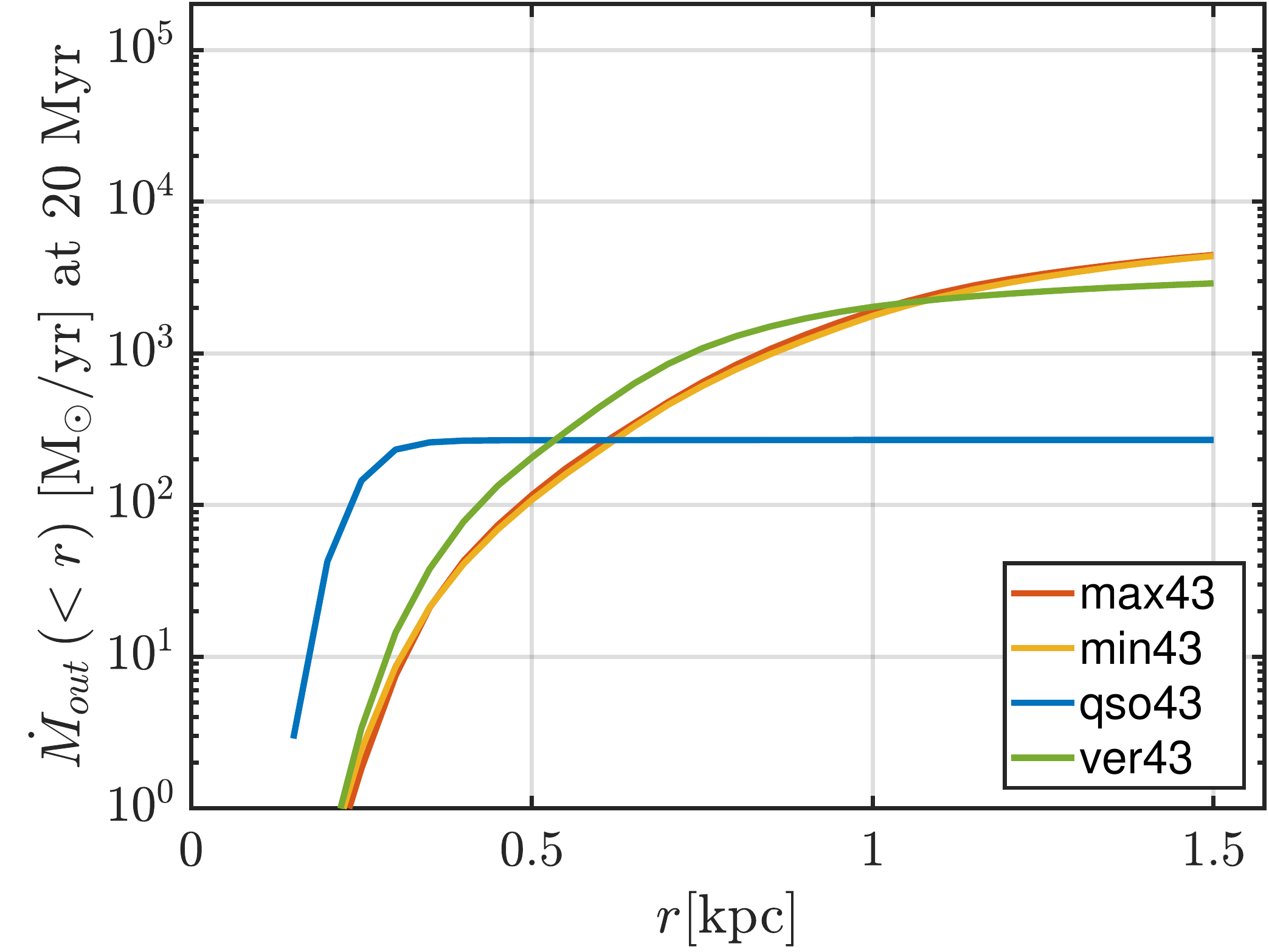}
  \includegraphics[width=\columnwidth]{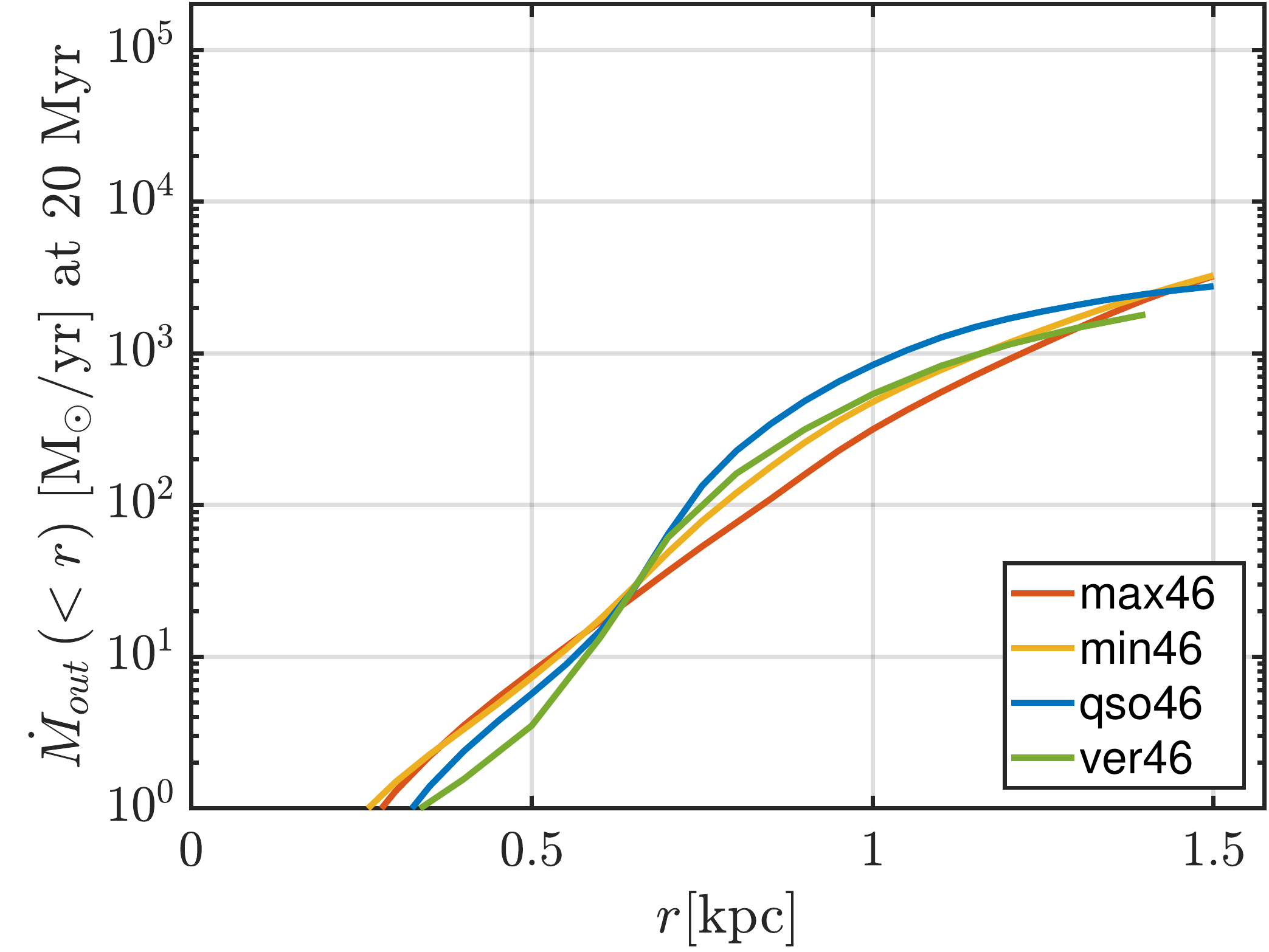}
 \caption{Cumulative profiles of the mass outflow rate (in solar masses per year), for both the p43 runs (left column) and the p46 runs (right column). At low power, all jets have very similar effects, but run qso43 is incapable of generating large-scale outflows. At high power, min46 and max46 can generate outflows up to $50000$~M$_\odot$/yr, while ver46 and qso46 have approximately the same outflow rate past $~250$~pc.
}\label{fig:flux}
\end{figure*}

Fig.~\ref{fig:flux} shows cumulative mass outflow profiles, spherically-averaged, at both $t_\mathrm{on}$ (top panels) and the end of the simulation. For the p43 runs (left column) the jets are capable of powering outflows of several thousand solar masses per year. The min43 and max43 jets continue powering outflows throughout the whole disk radius. Ver43 briefly overtakes them around $0.5$~kpc, only to approximately saturate around $1$~kpc (similar to the interaction radius we observed in Figure \ref{fig:tempfo}) around $3000$~M$_\odot$/yr. The run still features some outflows outside the disc plane past that radius, but the mass is negligible. Run qso43 exhibits a total outflow rate of ~M$_\odot$/yr.
The outflow is limited to the central $200$~pc, the size of its own central bubble. By the end of the simulation, gas is still flowing out and the outflow rate of the jets decreases only slightly. Some expansion is still ongoing, since the interaction radius of the AGN in the disc
in ver43 has increased a little.

The outflow rates at $t_{on}$ for the min and max jets are larger by almost a factor of ten, while ver46 shows again very little outflow past $1$~kpc. The qso46 run shows the highest outflow rate of all runs at about $500$~pc (most of the outflow mass being in its dense shell), and saturates near $10^4$~M$_\odot$/yr, about the same as ver46. At the end of the simulation, outflow rates decrease substantially, in part because the AGN have had the time to fully interact with the disc, and most of the outflowing mass is now located at larger radii. Run ver46 shows still very low mass outflow rates past $0.5$~kpc, and saturates at about $200$~M$_\odot$/yr, while all other runs end up at about $3000$.

We note that the outflows at $t_{on}$ are in agreement with the rate of fast outflows measured by \citet{tombesi_ufos_2015} of about $800$~M$_\odot$/yr. Most of the radiative runs in B+17 yielded similar predictions. This shows that both AGN modes couple efficiently with the ISM.

\begin{figure*}
  \includegraphics[width=\textwidth]{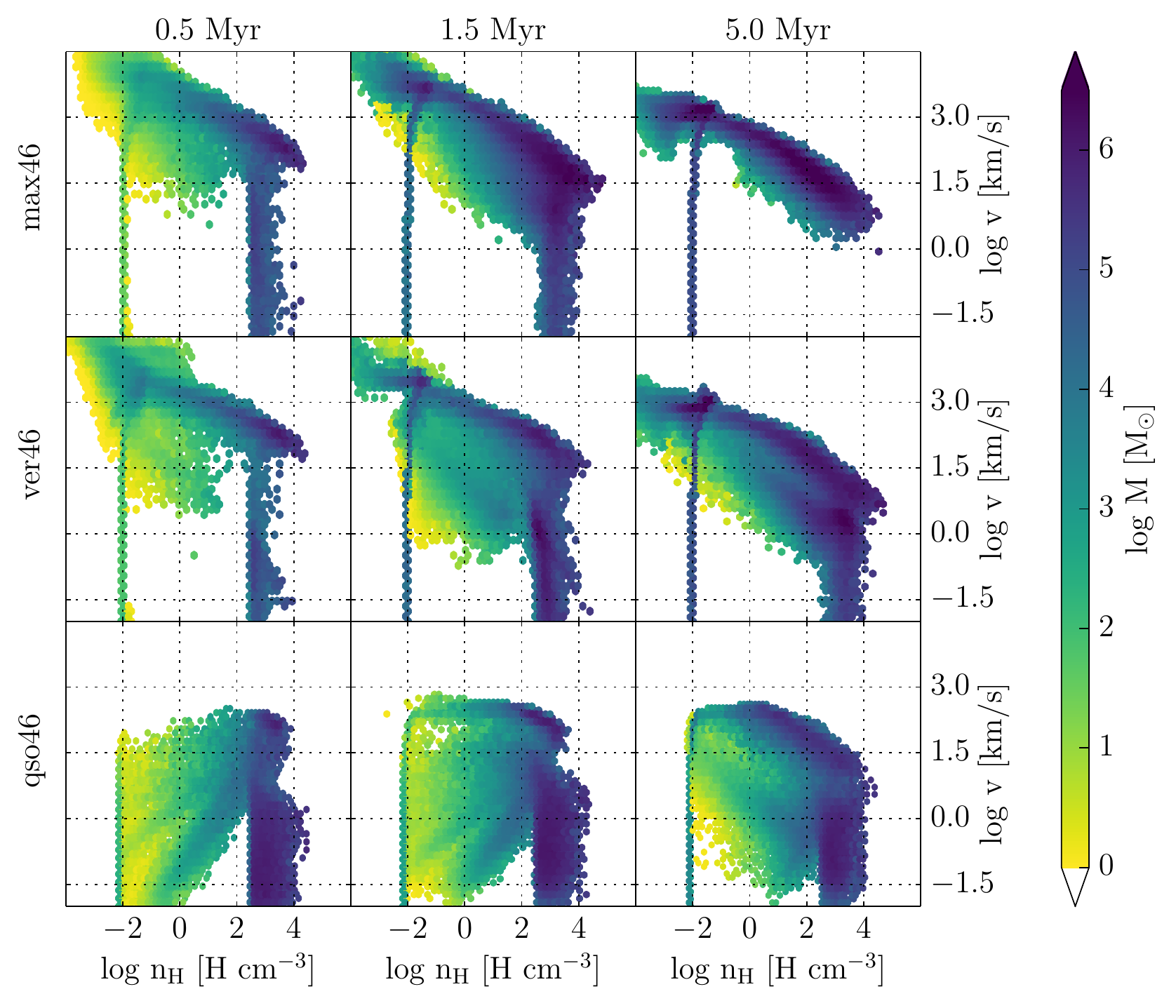}
\caption{Mass-weighted velocity versus density for the p46 simulations at 0.5, 1.5, and 5~Myr. The points are colored by the total mass within each histogram. The max46 simulation manages to accelerate most mass to high velocities, whereas the qso46 and vert46 simulations show similar velocities for gas densities above $10^{-4}$~cm$^{-3}$.
The vertical line at $10^{-2}$~H~cm$^{-3}$ arises from the low density hot phase gas right at the edge of the galaxy disc.}
  \label{fig:phase}
\end{figure*}

Figure \ref{fig:phase} shows 2D phase plots of the gas number density as a function of velocity for the high power runs max46 at $0.5$, $1.5$ and $5$~Myr. We omit the phase diagram of the min46 simulation as it is very similar to the max46 run.

A common feature in all panels is a narrow vertical line at $0.01$~cm$^{-3}$. That corresponds to the hot background density (in B+17 that gas is also shown to be at the background temperature) and corresponds to small amounts of hot ambient gas right at the edge of the disc that is pushed away from the disc by the AGN outflows.

As observed by B+17 for the case of radiative feedback, most of the gas shows an anti-correlation with density (i.e. the most populated regions of the diagrams show negative slopes), as radiation was observed to heat and rarefy the gas before accelerating it. This is  true in the case of jets as well, as the jets strip the clumps of their outer layers before reaching the cores.

One noteworthy exception to this anti-correlation trend is visible in run qso46 at very early times when the photons are still trapped within the initial overdensity. While the outflow is still confined within, the shock-compressed dense gas ($\gtrsim10^3$~cm$^{-3}$) moves at higher velocity ($\gtrsim100$~km/s) than lower density gas. Once the outflow escapes the central clump (around $1.5$~Myr) a tail at higher velocities but lower densities develops, before the AGN is finally switched off and the outflow velocities start to decline again. 

The radiative runs create much more intermediate density gas ($0.01-10$~cm$^{-3}$) than jets, as show by the larger population in the low-density, low-velocity region of the diagram. In the jet cases, the low-density gas comes 
instead directly from the jet beam or the initial shocks, thus have velocities in the thousands of km/s range (by construction) but consist primarily of material injected with the jets. This confirm that jets have a high momentum escape fraction (see also Section \ref{sub:dpdt}). Despite that, jets can accelerate dense gas to higher velocities than in the qso run. The outflows reach, at their peak, velocities exceeding $1000$~km/s, even for densities higher than $100$~cm$^{-3}$. In the absence of cooling, these results show that jets are able to accelerate some of the densest central clumps for a considerable fraction of a Myr before the clumps heat up and expand. 

The heat transfer may  be higher in the presence of efficient thermal conduction. Following \citet{cowie_evaporation_1977}, we can estimate the evaporation time $t_{evap}$ due to thermal conduction for the gas clump shown in Figure \ref{fig:IC}, (i.e. a clump of $100$~pc in size, and with a typical number density of $1000$~cm$^{-3}$). We find that $t_{evap}\sim30$~Myr for a hot gas phase of $10^8$~K. 
Thus, even when highly efficient, (like in the runs in which a very hot gas phase percolates through the disc, max46 and min46),
thermal conduction is unlikely to counteracting cooling (see Section \ref{sec:fb}) or heat the accelerating gas.

All the jet simulations have fast-outflowing gas at intermediate densities (becoming denser but slower with time, as extra gas is swept up) that may condense and give rise to observed fast and cold outflows. Direct acceleration of the dense gas is most effective in the max46 (and min46) case, as the jet spends more time in contact with the dense clumps in the disc. Jet/ISM interaction in a similar context have been observed, for instance, by \citet{morganti_youngjet_2015}, who traced the kinematic of the ISM affected by jets powering fast outflows in young radio sources. Their HI and molecular emission favor a clumpy ISM model.

\subsection{Mechanical coupling to the cold ISM}\label{sub:dpdt}

The efficiency of the coupling between the AGN and ISM can be quantified by the momentum deposited by the jet/quasar into the gas. We thus calculate the \emph{mechanical advantage} of the AGN, defined as the time derivative of the total momentum of the gas, in units of $P_\mathrm{AGN}/c$ (i.e. the constant $10^{43}$ or $10^{46}$~erg/s divided by the speed of light). A mechanical advantage above unity generally occurs when the AGN outflow is energy-conserving and the momentum of the gas builds up from thermal pressure gradients acting on the gas. Values much larger than unity have been measured in galactic outflows (up to several hundreds for central, molecular, ultra-fast outflows, e.g. \citealp{tombesi_ufos_2015}).

Such high kinetic coupling can be achieved in an energy-conserving regime. Energy can be stored in the shock-heated gas, which will then expand supersonically;  for radiative feedback, multiple photon scattering can lead to a similar result, as shown in B+17.

\begin{figure}
  \includegraphics[width=\columnwidth]{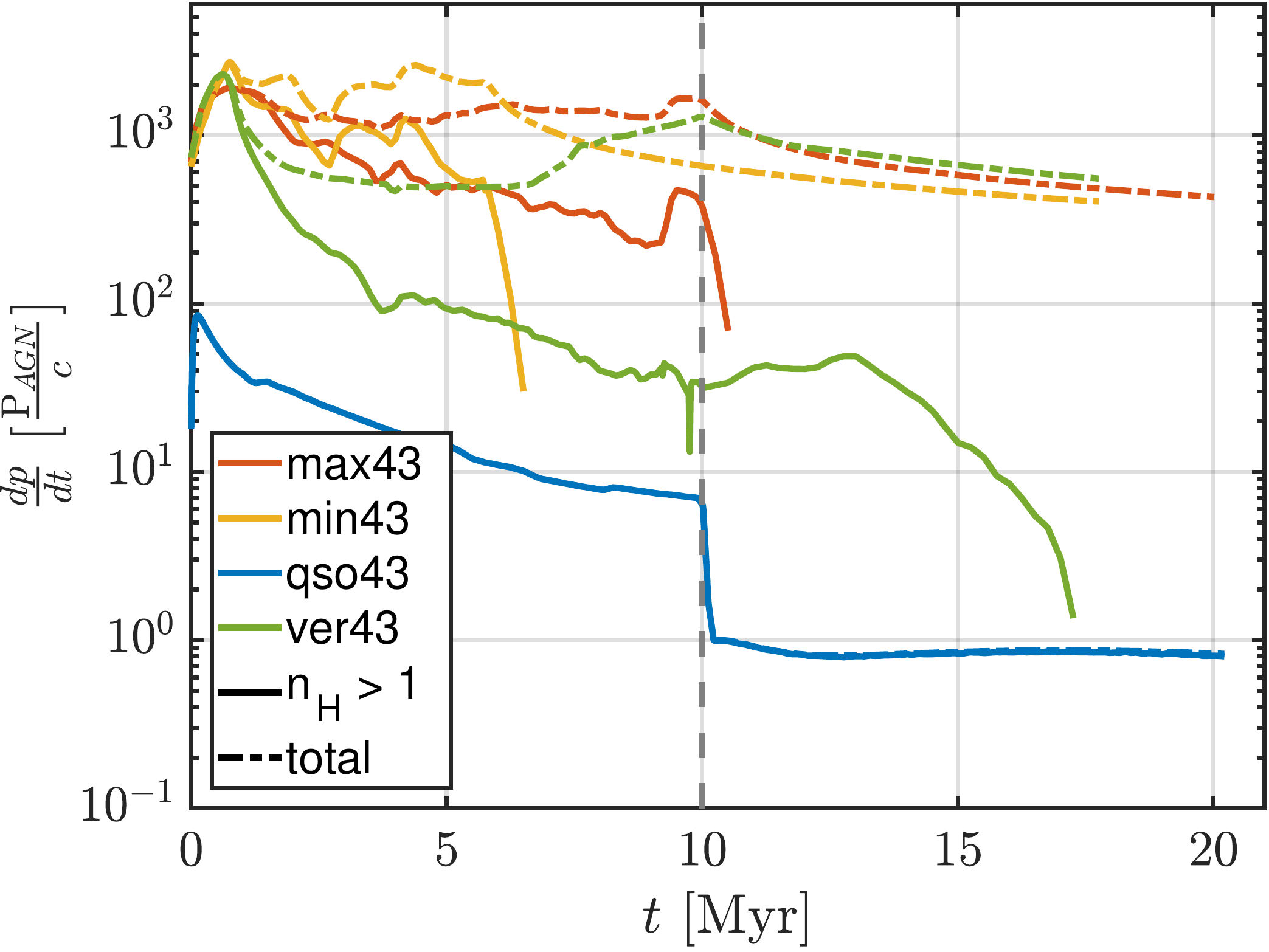}
  \includegraphics[width=\columnwidth]{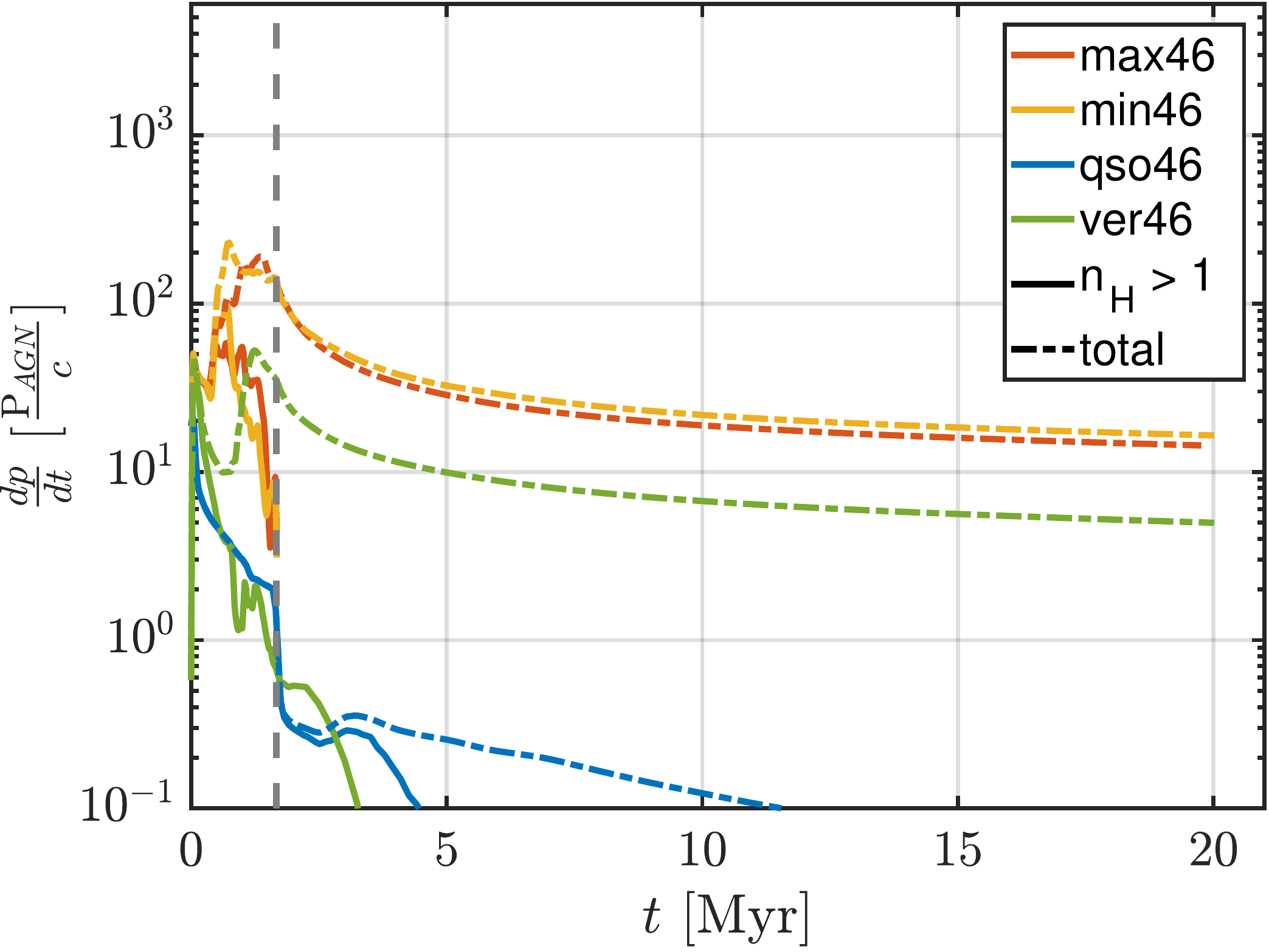}
  \caption{Time evolution of the mechanical advantage for all gas (dash-dot lines) and for gas denser than $1$~H/cc (solid lines). The mechanical advantage is higher at low power. The runs with the highest mechanical advantage are always the min/max runs.}  \label{fig:dpdt}
\end{figure}

Our simulations give a dynamical view of the mechanical coupling, and predict whether either radiative and jet AGN feedback can explain the observed outflows. In Figure \ref{fig:dpdt} we plot the evolution of the mechanical advantage, for the p43 simulations (top) and p46 simulations (bottom).

We show both the total mechanical advantage (dash-dot lines) and the mechanical advantage for the dense ISM (solid lines), where we only consider the momentum of the gas higher than a density of $1$~cm$^{-3}$. This is done to exclude the momentum of the jet material itself and of the hot gas swept-up by the expanding cavities. These can in fact escape the galaxy without significantly accelerating the cold ISM.


The mechanical advantage is highest in the low-power case (similar to \citealp{wagner_driving_2012} and \citealp{bieri_outflows_2017}). In the initial phase within the central clump, the mechanical advantage is maximal, getting up to about $3000$ for all jets and 
$100$ for the qso run; the high power case shows more moderate values of about $200$ decreasing to $10$ (min46/max46) or $10$ decreasing to $1$ and below (ver46/qso46). These values are in agreement with \citet{wagner_erratum_2011} studying the interaction of jets with a fractal ISM of elliptical galaxies, although the simulation time was less than a Myr. 

At low power, the mechanical advantage decreases as the outflows get out the central overdensity, and
increases again when new dense gas is encountered. This happens much more frequently for the min43 and min46 cases, which can interact with the disc material most effectively, for purely geometrical reason (in other words, the effective volume covering factor in jet beam is larger for them).

The mechanical advantage decreases sharply to zero as the jet is switched off (max43) or the beam is deflected outside the disc plane (min43), since the beam can no longer impinge on the cold disc gas. 
The ver43 jet gets out of the disc plane as soon as it breaks out the central overdensity. While its total momentum coupling is still in line with the one of the other jets (in the several hundreds), the momentum coupling with the dense-gas drops to about $50$ in a few Myr, where it remains for several Myr, even after $t_\mathrm{on}$. This is a sign that the cavities continue to interact with the dense material of the disc, and with the warm, dense gas phase that was lifted off the disc plane, even for longer than the min/max jets do.
Run qso43 exhibits a lower mechanical advantage than the runs with jets, from $100$ decreasing down to about $10$ at $t_\mathrm{on}$, dropping sharply when the AGN switches off. The reason why radiation shows, in general, a less efficient momentum coupling between the AGN and ISM is discussed in detail in B+17. Briefly, IR radiation is mostly responsible for transferring momentum to the gas via multiple scattering on dust in the dense clouds. However, the non-uniform structure of the ISM and the subsequent building of low density channels by the photons as well as the creation of a cavity in the central region of the galaxy, causes the mechanical advantage to quickly decline as the central gas cloud expands and breaks up. 

Because the radiation-driven wind in qso43 never manages to break out of the central clump, its momentum coupling only affects the dense gas (i.e. the solid line and the dash-dot line are indistinguishable) of the swept-up shell.

The mechanical advantage is more moderate for high power AGN in which the explosive evolution results in the jets and the photons to propagate quickly beyond the gaseous ISM in less than one Myr (in agreement with the findings of \citealp{mukherjee_relativisticdynamics_2016}). Thus, the radiation and the jets do not encounter as much gas to couple with.  Min46 and max46 undergo a very similar evolution. This is because at high power they can interact with the disc globally (we saw in Figure \ref{fig:dens46} how they pervade the disc entirely). Their dense-gas mechanical advantage shows a sharp drop right before the jets switch off (after which gas starts to decelerate), meaning that the lifetime we estimated roughly coincides with the time of their active interaction with the disc. This just implies that even if $t_{on}$ was longer, the evolution of the mechanical advantage wouldn't change by a large extent.

The total mechanical coupling decreases only slightly after that, which is a sign that the hot outflows and bubbles still accelerate and expand in the hot ISM. Note that the total mechanical coupling is dominated by low density gas, showing that we are just observing the kinematics of the hot outflows, i.e. the physical processes associated with classical \emph{radio-mode} feedback. These are usually more important in larger halos, cluster-central galaxies, in which the energy stored in the cavities can gradually heat the surrounding gas. In line with this, run qso46 has a much lower total momentum coupling, almost always below $1$. 

Also remarkable is the very similar evolution of runs ver46 and qso46 for the mechanical advantage with respect to the dense gas, which has decreased to a value of $1$ or $2$ by the time the AGN is switched off.
This shows that both radiation and high power jets, when allowed to avoid direct interaction with clumps, take very soon the path of least resistance perpendicular to the disc, thus accelerating gas only mildly and for a short time.



\section{Small-scale inflows and accretion} \label{sec:inflows}

In all our runs we observe mass inflows, on scales spanning from a few to several hundred pc, triggered by hydrodynamics and thermodynamics only. These inflows can be seen as a generalisation of the backflows studied in C+17, which considered only jets and did not include a gaseous disc.
The physical driving processes of inflows varies in the different AGN models, and the classical backflow description does not apply to all of them. In this section we describe the inflows, quantify their rates, and investigate how they affect SMBH duty cycles (feeding/feedback cycle).

\begin{figure*}
\includegraphics[width=\columnwidth]{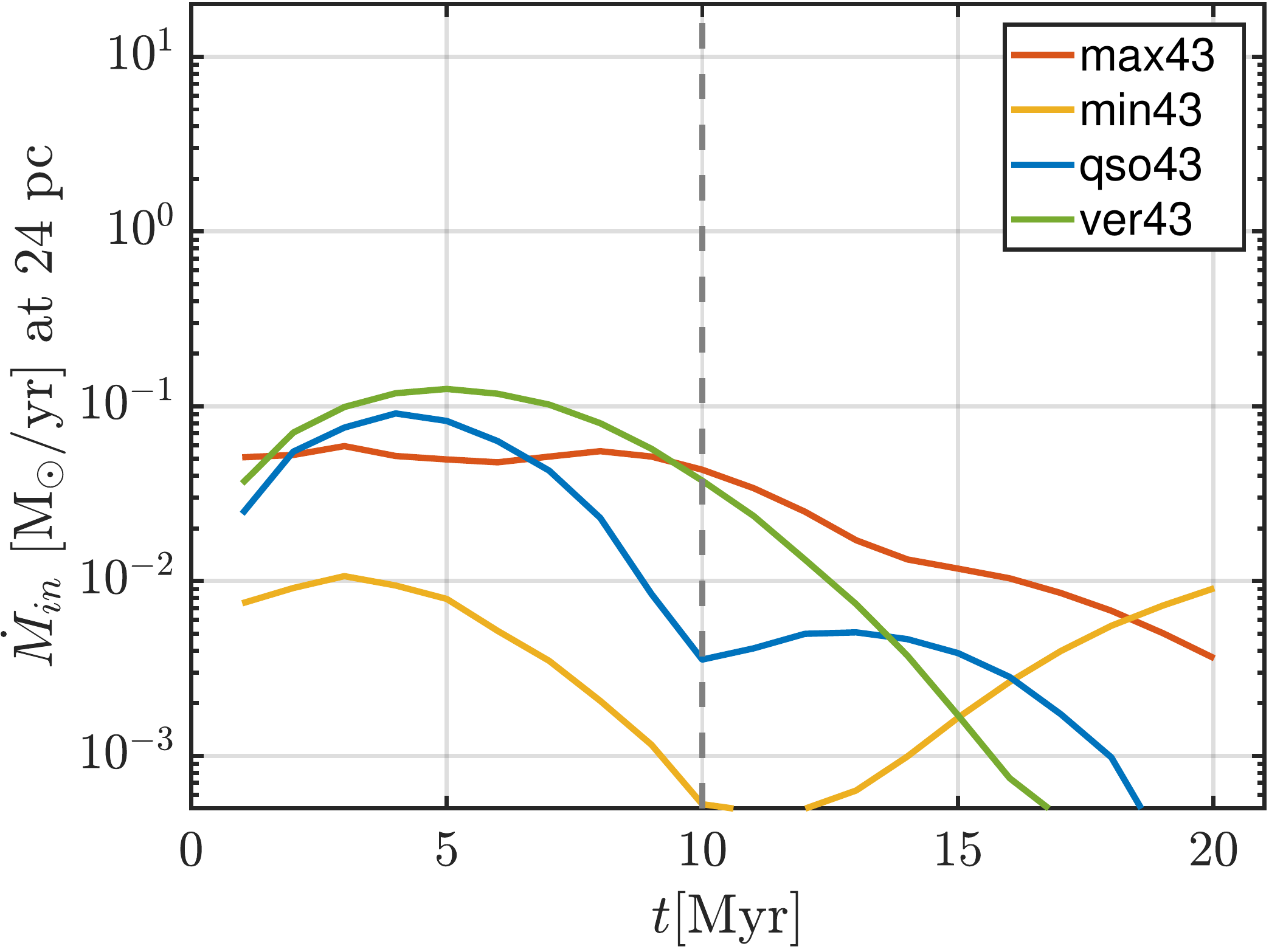}
\includegraphics[width=\columnwidth]{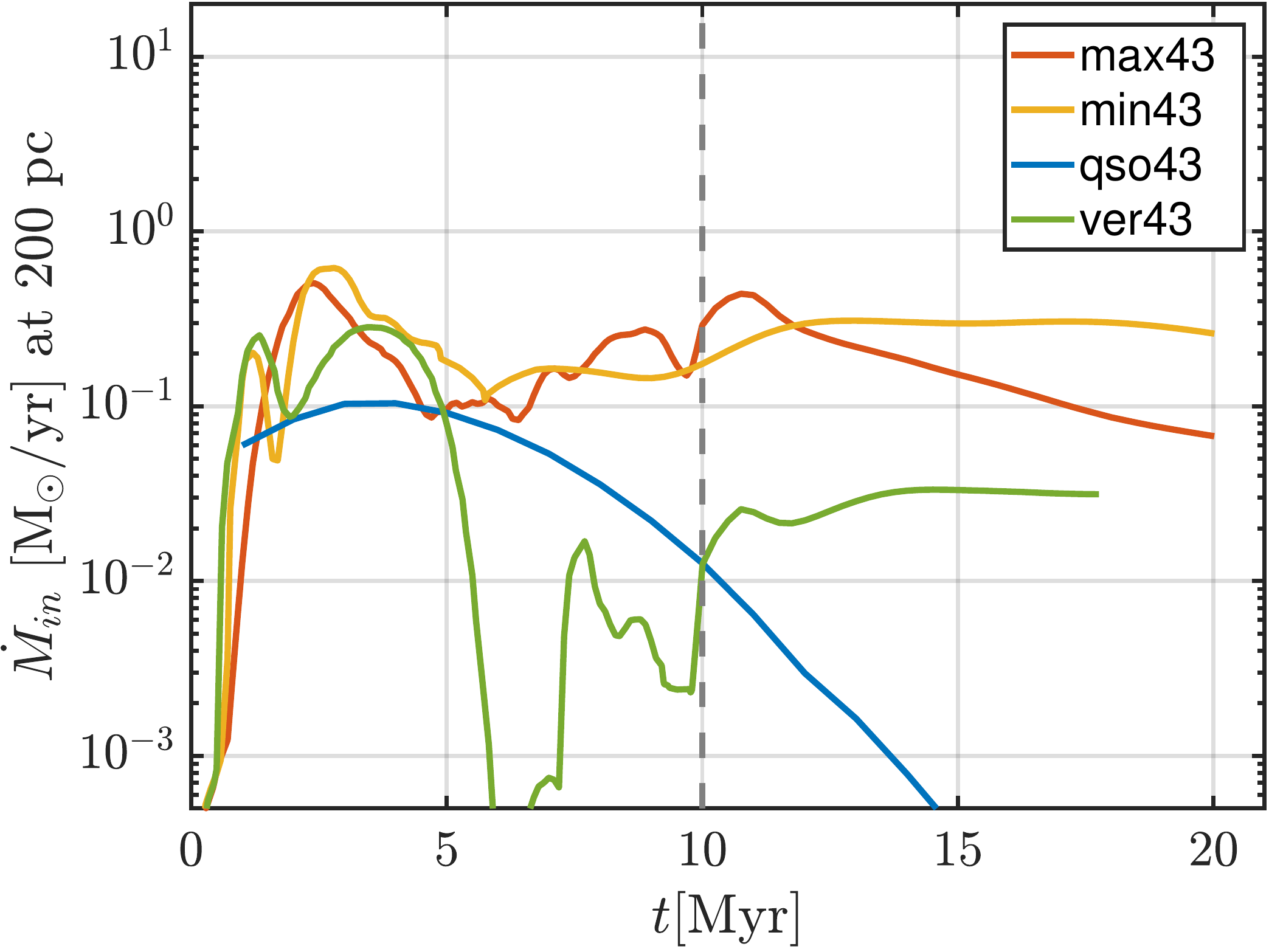}
\includegraphics[width=\columnwidth]{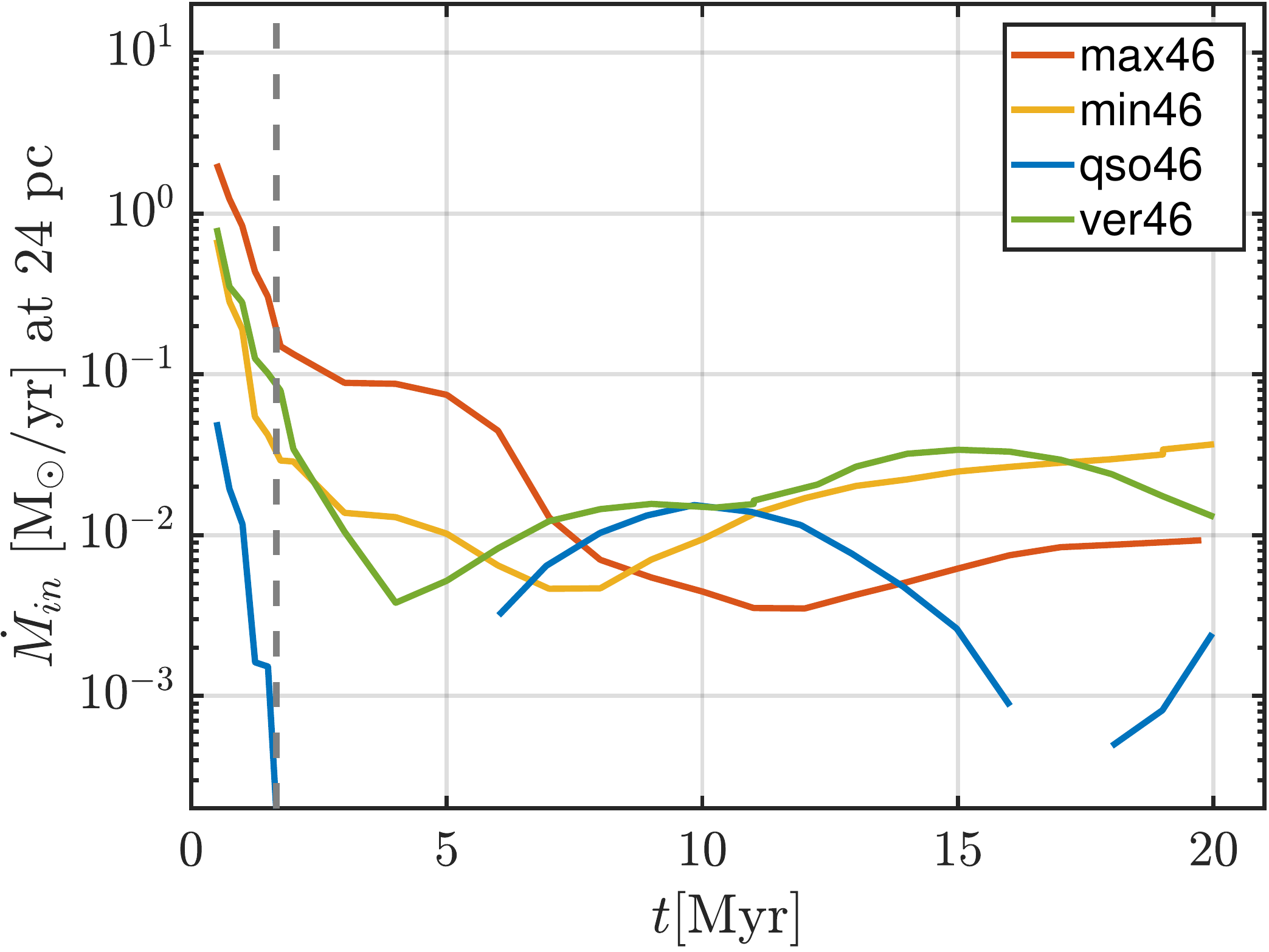}
\includegraphics[width=\columnwidth]{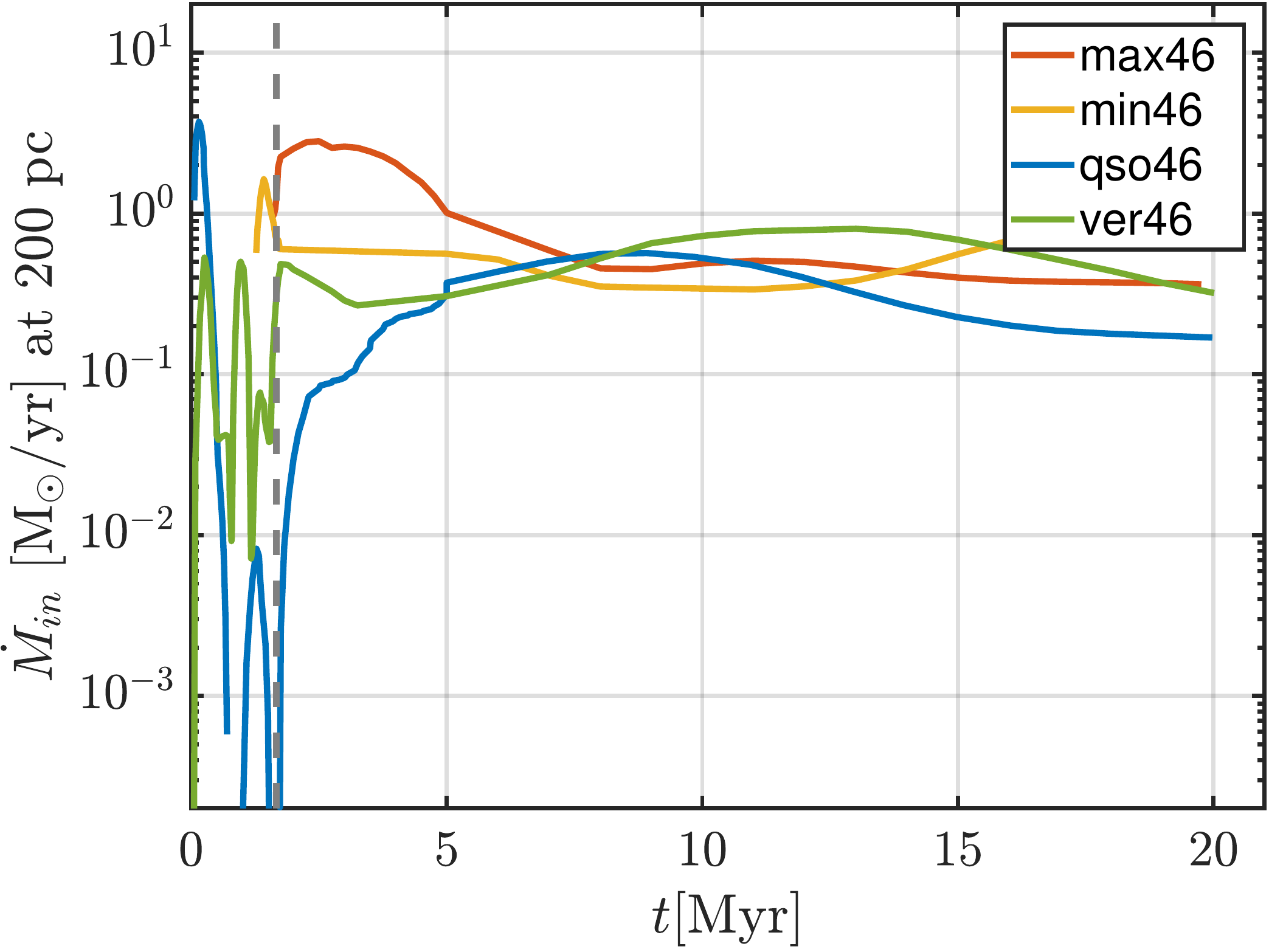}
\caption{Small-scale ($24$~pc, left column) and large-scale ($200$~pc, right column) mass inflow rate for both the p43 (top) and p46 runs (bottom). A 5-point moving-average smoothing is applied for visual clarity. Small-scale inflows range between $0.01-0.1$~M$_\odot$/yr, with an initial peak associated with the initial shock, then a slow (p43) or rapid (p46) fall, 
but the inflows continue for several tens of Myr. On large scales inflows are more significant ($0.1-1$~M$_\odot$/yr). }
\label{fig:inflowmass}
\end{figure*}

Figure \ref{fig:inflowmass} show the gas inflow through spherical surfaces of radius $200$ and $24$~pc. Mass inflow rates are obtained by selecting only the gas with negative (i.e. pointing towards the origin) radial velocity. 
Inflows are not necessarily present all the time, and when they are not the lines in the figure are broken. However, when present, inflows appear as continuous and coherent inward motions of dense gas, and are observed in all runs within $24$ to $500$~pc distance. Below $24$~pc the flux calculation is just affected by poor statistical sampling, given our $5.8$~pc resolution, and by the fact that the center is not maximally refined when no cold gas is present. In the quasi-stationary set-up of the simulations, inflows are generated by backflows as in C+17, but also reflected pressure waves or reverse shocks in the regions affected by the AGN outflows (see Section \ref{sub:feeding}). Thus, there can be no inflow from scales not yet affected by the AGN, e.g., during the first Myr in run qso43.


These inflows may appear to carry little mass, yet in the context of feeding the SMBH and sustaining an AGN, they  are very significant: a mass inflow rate of $\dot{M} = 0.01$~M$_\odot$/yr corresponds to an AGN power of $P_\mathrm{AGN} = \varepsilon \dot{M} c^2 = 5.7\times10^{43}$~erg/s for $\varepsilon=0.1$, even larger than the initial p43. This corresponds to $L/L_{\rm Edd}=0.44 \dot{M}_{-2} M^{-1}_\mathrm{BH,6}$, where the inflow rate is expressed in units of $10^{-2}$~M$_\odot$/yr and the SMBH mass in units of $10^6$~M$_\odot$. For a small SMBH, e.g., $10^6$~M$_\odot$ , this gives $L=0.44\times L_{\rm Edd}$, and according to the ``accretion paradigm" discussed in the Introduction, would produce a radiatively efficient AGN, i.e. an analogue of qso43. For a large BH e.g., $10^9$~M$_\odot$, this gives $L=4.4\times 10^{-4}\times L_{\rm Edd}$, and the result would be a jet similar to the p43 we have simulated. In both cases, this mechanism could lead to self-regulation of AGN through a cycle of feedback and feeding.

\subsection{Inflow scales and feeding of the Black Hole}\label{sub:feeding}

As an estimate of the smallest scales the inflow can reach, we investigate the distribution of the impact parameter $b$ of the inflowing gas. Given the absence of gravity and rotation in our simulations, $b$ is a scale determined only by the hydrodynamics induced by the AGN. For instance, \citet{antonuccio-delogu_feeding_2010} and C+17 have shown how the evolution of the \emph{backflows} is determined by thermodynamics alone and down to scales of a few tens of parsecs, even in the presence of gravity and radiative cooling processes. Our inflows are not limited to the disc plane, so we use the three-dimensional expression for $b$, which can be directly estimated from the position $r$ and the radial-to-total velocity ratio of the inflows:
\begin{equation}
b = r \sqrt{1-\frac{v_\mathrm{r}^2}{v^2}} \quad .
\end{equation} 
The impact parameter $b$ contains information on position, direction and kinetic energy of the inflow.

\begin{figure}
\includegraphics[width=\columnwidth]{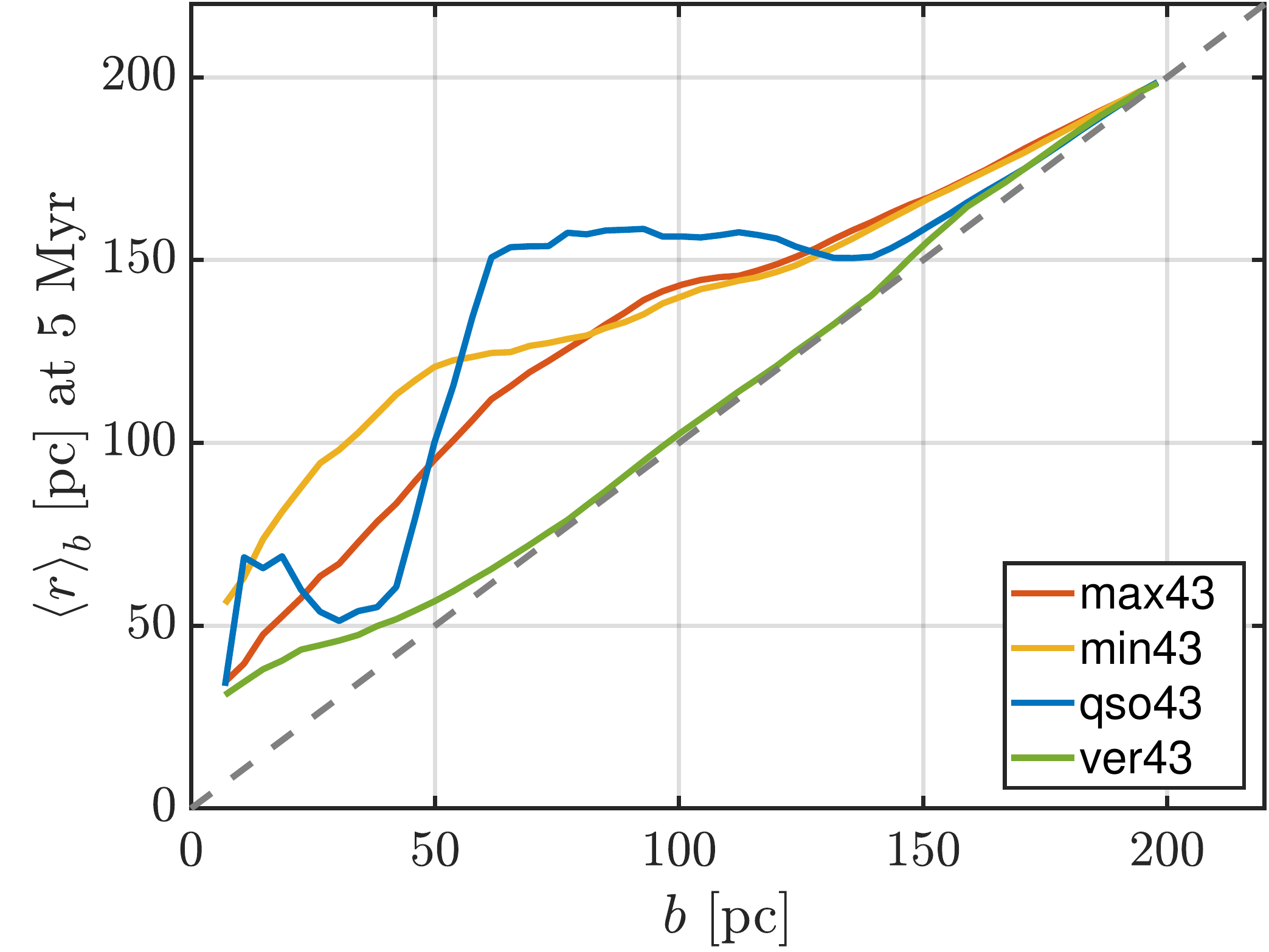}
\includegraphics[width=\columnwidth]{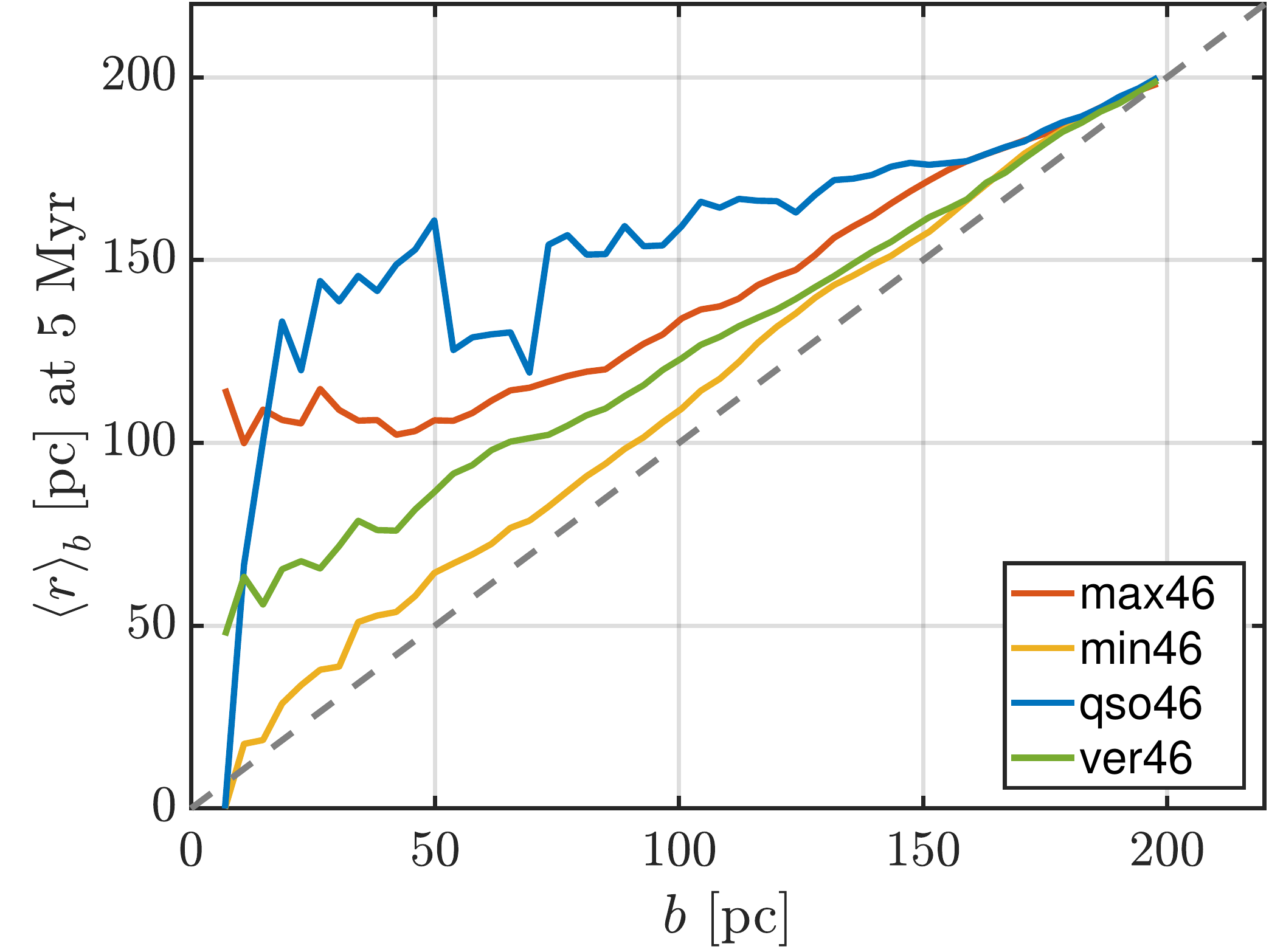}
\caption{3D impact parameter $b$ of the inflow, against its ``provenance radius'' $\left\langle r\right\rangle_b$, i.e. the average gas radius in bins of b. The values are calculated at $5$~Myr time, for all inflowing gas within $200$~pc from the center,as in Figure \ref{fig:inflowmass}. The dash-dot line marks $r=b$. }\label{fig:b}.
\end{figure}

In Figure \ref{fig:b}, we calculate $b$ for all gas within $200$~pc, and plot it against the average gas position per bin of $b$. This quantity indicated by $\left\langle r\right\rangle_b$ is the ``provenance'' radius of the inflow gas, 
a measure of both the spatial extent and the speed of the radial migration: fast-moving gas is expected to quickly reach its $b$ value, while slower inflows will show larger deviations from $r=b$ line.


The largest deviations from that line are observed in run qso43, where large amounts of gas within $150$~pc is moving inwards, down to about $50$~pc (central plateau in the curve), with an additional migration represented by a peak from $75$ to $25$~pc. 
The regions with very low impact parameters in the qso runs are due to inflow caused by reverse-shocks or outflows reflecting from the inner side of the quasi-spherical initial cavity, thus introducing very little deviation from the radial direction. 

The ver43 lies almost on the $r=b$ line, which is a sign that most of its fast inflows have already reached their radii comparable to their impact parameter, except within the inner $\sim40$~pc, where gas is still migrating inwards. The min43 and max43 runs are in an intermediate position, except for gas between about $15$~pc to $50$~pc, where min43 shows the highest inward migration. At this time, the jet beam of min43 has already been deflected outside the disc plane, and it is triggering backflows from outside.
For the p46 family of runs, qso46 still shows a sign of a large inwards migration from a radius of $150$~pc. Max46 exhibits a plateau at $\left\langle r\right\rangle_\mathrm{b}=100$~pc, while run ver46 shows no significant plateau. Finally, the min46 run is always very close to the $r=b$ line.

\begin{figure}
\includegraphics[width=\columnwidth]{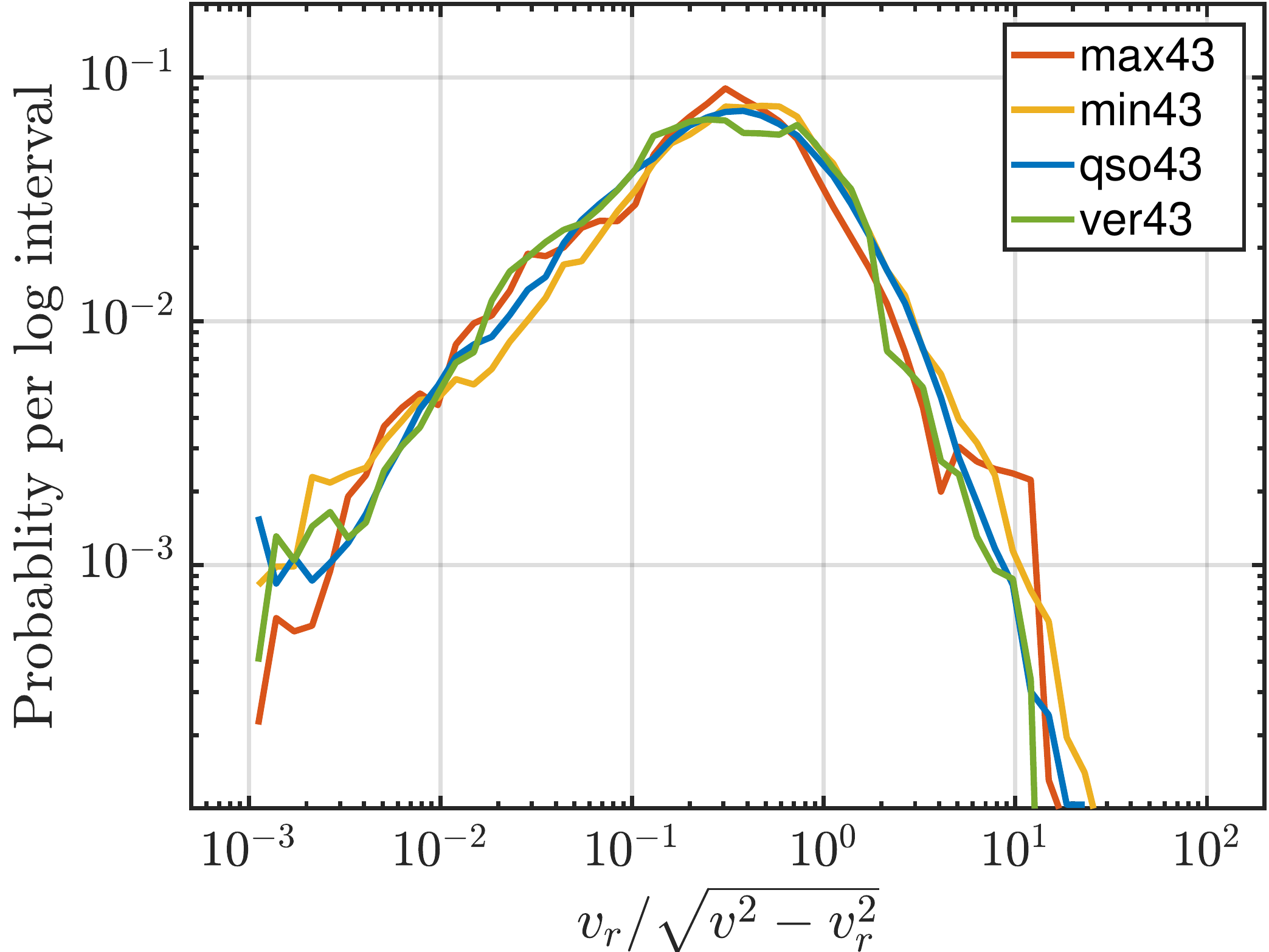}
\includegraphics[width=\columnwidth]{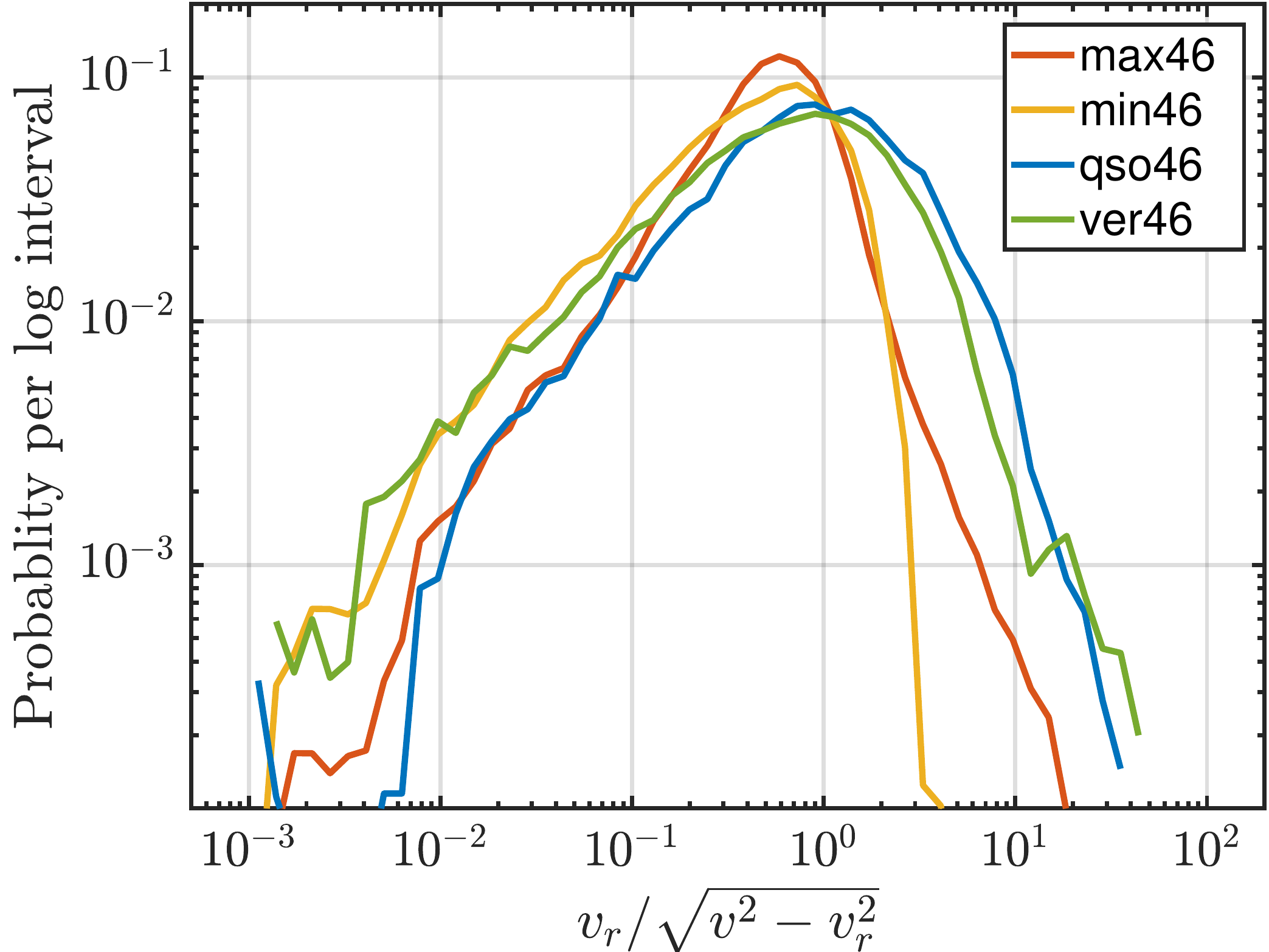}
\caption{Probability distribution of the radial-to-tangential (in 3D) velocity ratio at $5$~Myr. All distributions peak at values slightly below unity, showing a balance of radial and tangential motions in the inflows.}\label{fig:vratio}.
\end{figure}

Figure \ref{fig:vratio} shows the (volume-weighted) probability distribution of the ratio of radial-to-tangential velocity $v_\mathrm{r}/v_\theta$ (in 3D-spherical coordinates) for all inflowing gas within $200$~pc, again considering the snapshots at $5$~Myr. Gas with a dominant radial component has eccentric trajectories that bring it close to the BH, and make it more readily available for accretion. In all our simulations, about one percent of the gas has $v_r  \geq 10 v_\theta$, and about 10 per cent (about 40 in qso46 and ver46) has $v_r  \geq v_\theta$. The distributions are fairly similar for all runs, peaking around $0.3$ for the p43 series, $0.8$ for the p46 ones (slightly higher for qso46 and ver46). Overall, the plots show a balance of radial and tangential motions. 

\begin{figure}
\includegraphics[width=\columnwidth]{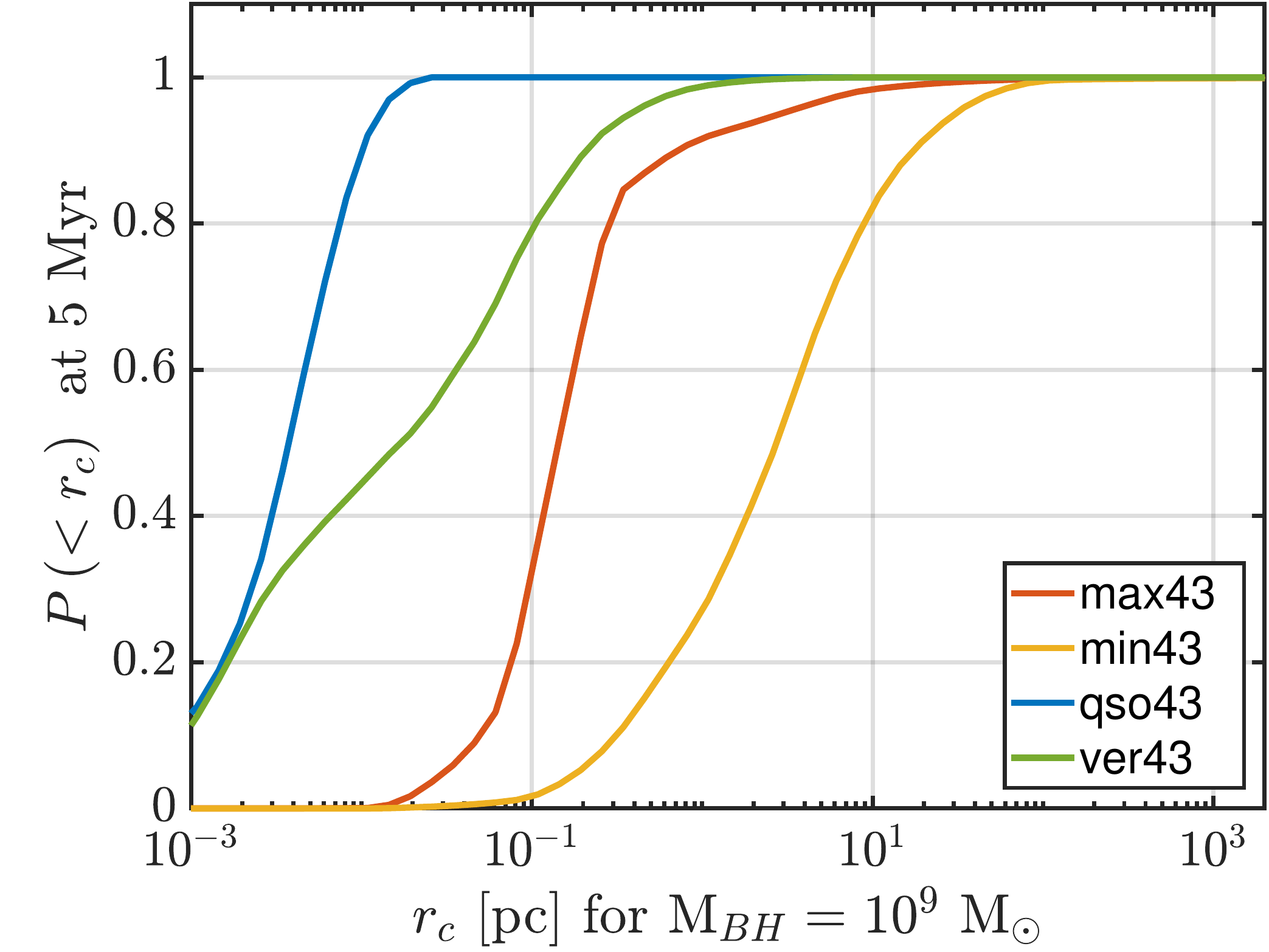}
\includegraphics[width=\columnwidth]{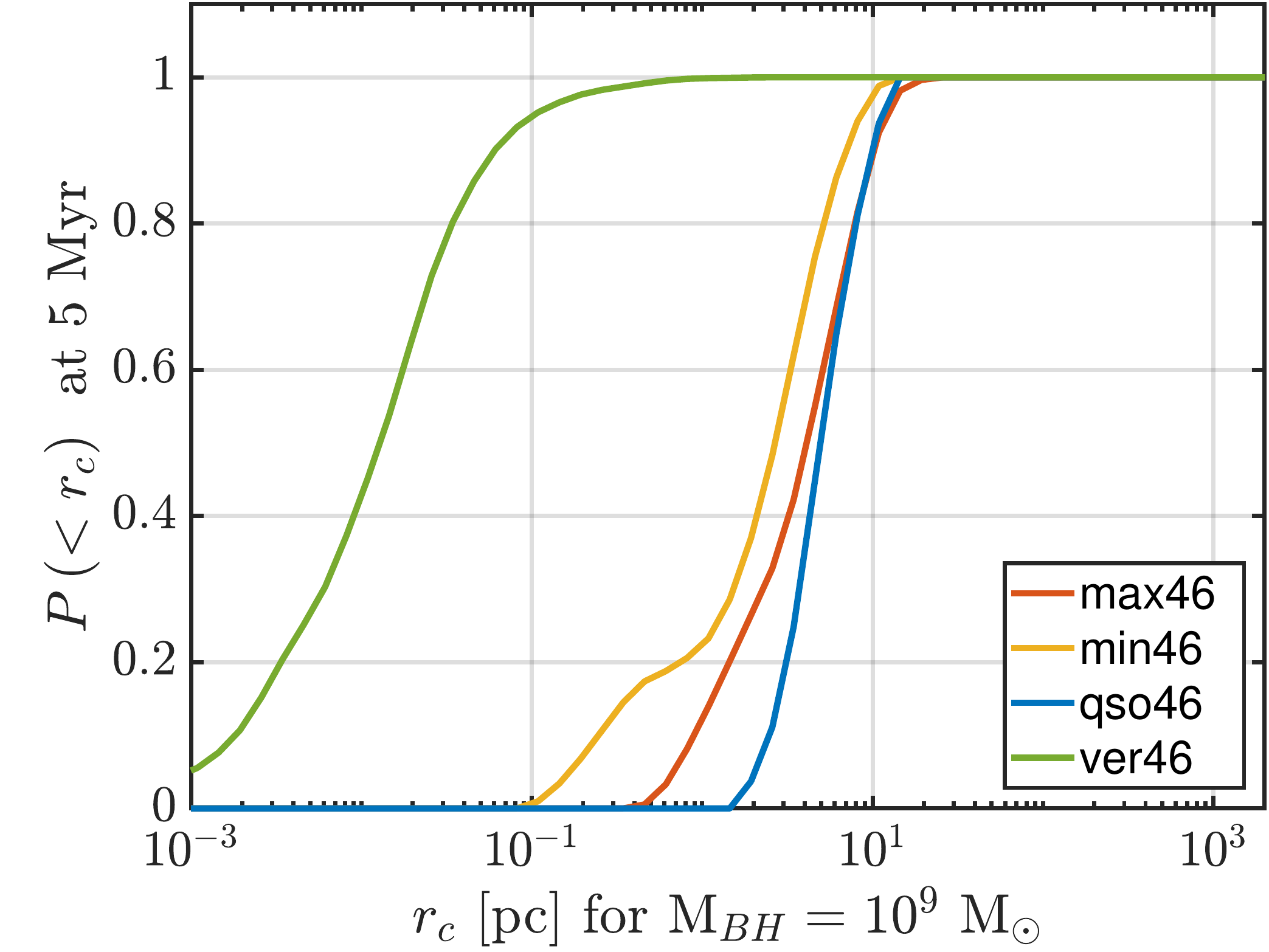}
\caption{Cumulative distribution function of the orbital radius $r_c$ (equation \ref{eq:orbit} in the text), evaluated at $5$~Myr for all the inflowing gas within $50$ pc, assuming a central mass of $10^9$~M$_\odot$. At high power, all runs but ver46 have more than $80\%$ of their gas  with $r_c$ between $1$ and $10$~pc. At low power $r_c$ is much smaller, especially in the qso and ver cases.}
\label{fig:orbit}
\end{figure}

So far we have considered the geometry of the flow, but in order to estimate how close the inflows would get to the central SMBH, we need to take into account the full angular momentum, $l$ (at the beginning of the simulations, $l=0$ since the gas is static; this would not be the case with gravity and rotationally-supported gas). Assuming conservation of specific angular momentum, we can estimate where the inflow would circularize at radius $r_c$ around a gravitational mass $\mu\left( r_\mathrm{c} \right)$. In our simulations only the disc gas  would contribute to $\mu$, but in a more realistic scenario $\mu$ will likely include just the central BH, and part of a disc-like stellar component --- but only when looking at scales of a few hundred parsecs or larger. While we observe spatially-coherent inflows up to about $500$~pc, this analysis is mostly interesting case for the inflows closest to the center. So this time we consider all the gas within $50$~pc. We then have:
\begin{equation} \label{eq:orbit}
\mu\left(r_\mathrm{c} \right) r_\mathrm{c} = \frac{l^2}{G}\;,
\end{equation}
where $G$ is the gravitational constant. The (mass-weighted) cumulative probability distribution for $r_c$. is shown in Figure \ref{fig:orbit}, where we have assumed $\mu=10^9$~M$_\odot=$~constant.   Runs ver43 and especially qso43 have very high inflow fractions for $r_c<0.1$~pc; at high power this is true only for the very early first inflow peaks, well within $t_\mathrm{on}$. At $5$~Myr all runs in p46, except ver46, have $r_\mathrm{c}$ between $1$ and $10$~pc; these are still penetrating inflows, but suggest that gas becomes more rotationally supported at late times. 


In the case a super-massive black hole of $10^8-10^9$~M$_\odot$ the central potential would be dominated by the SMBH, and in this limit we can compare $r_c$ to the Bondi radius, defined as the radius within which the BH gravity dominates over the hydrodynamics: $r_\mathrm{B} := 2GM_\mathrm{BH}/c^2_\mathrm{s}$, where $c_\mathrm{s}$ is the local speed of sound. In order to compute $c_\mathrm{s}$ we first select four concentric spheres around the center of the galaxy, with radii of $24$, $36$, $48$ and $60$~pc. We then mask the high-temperature gas coming directly from the AGN or the innermost shocks by imposing an upper temperature threshold of $10^{10}$~K and applying our ideal gas equation of state to the remaining gas, averaging on the selection. Spatial convergence of the measured $c_\mathrm{s}$ is always observed for increasing radius (usually already at $24$ or $36$~pc).  A large fraction of the volume around the BH is initially occupied by very hot shocked gas, so that $c_\mathrm{s}$ ranges from one to a few hundreds km/s.

\begin{figure}
\includegraphics[width=\columnwidth]{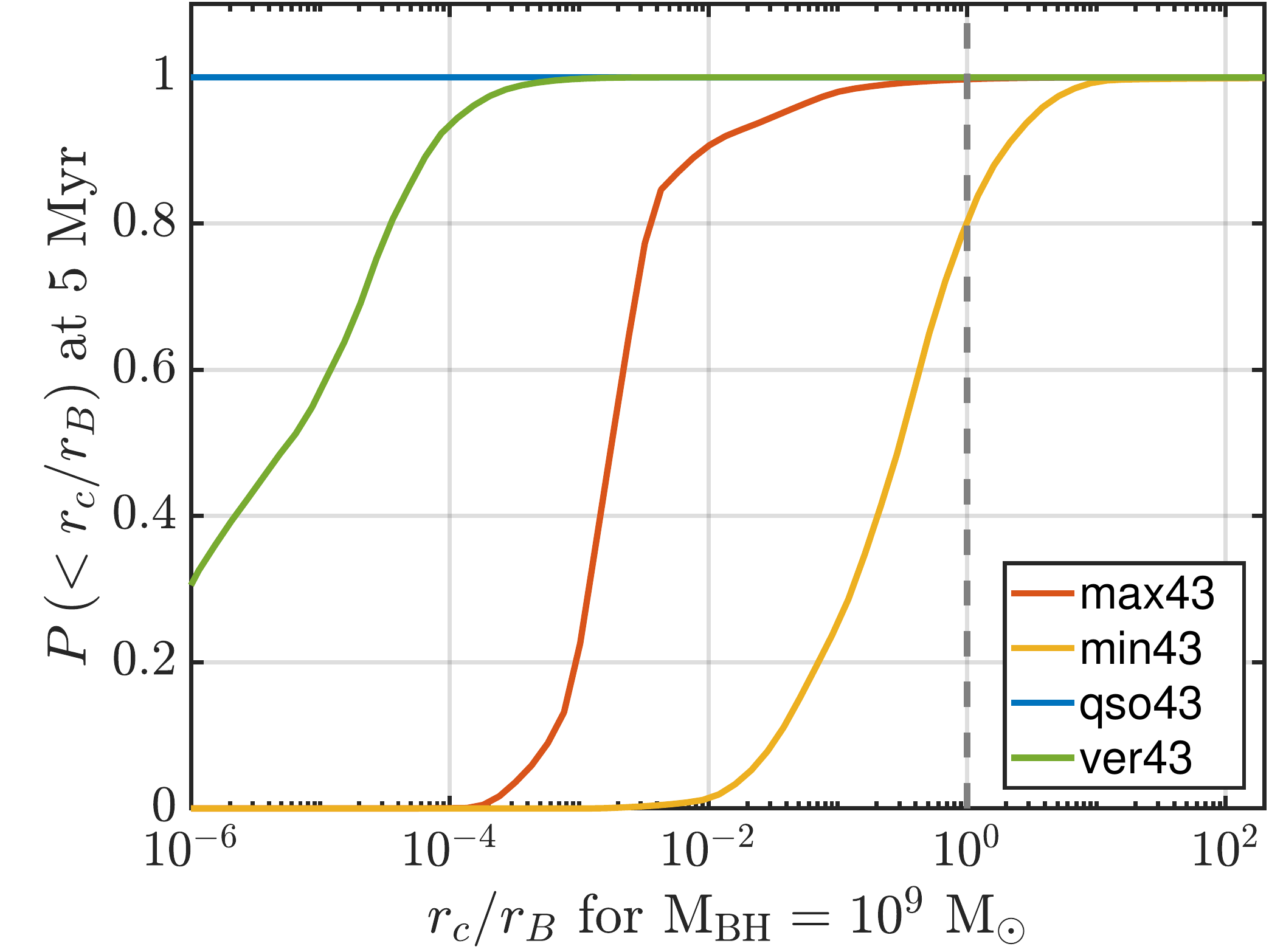}
\includegraphics[width=\columnwidth]{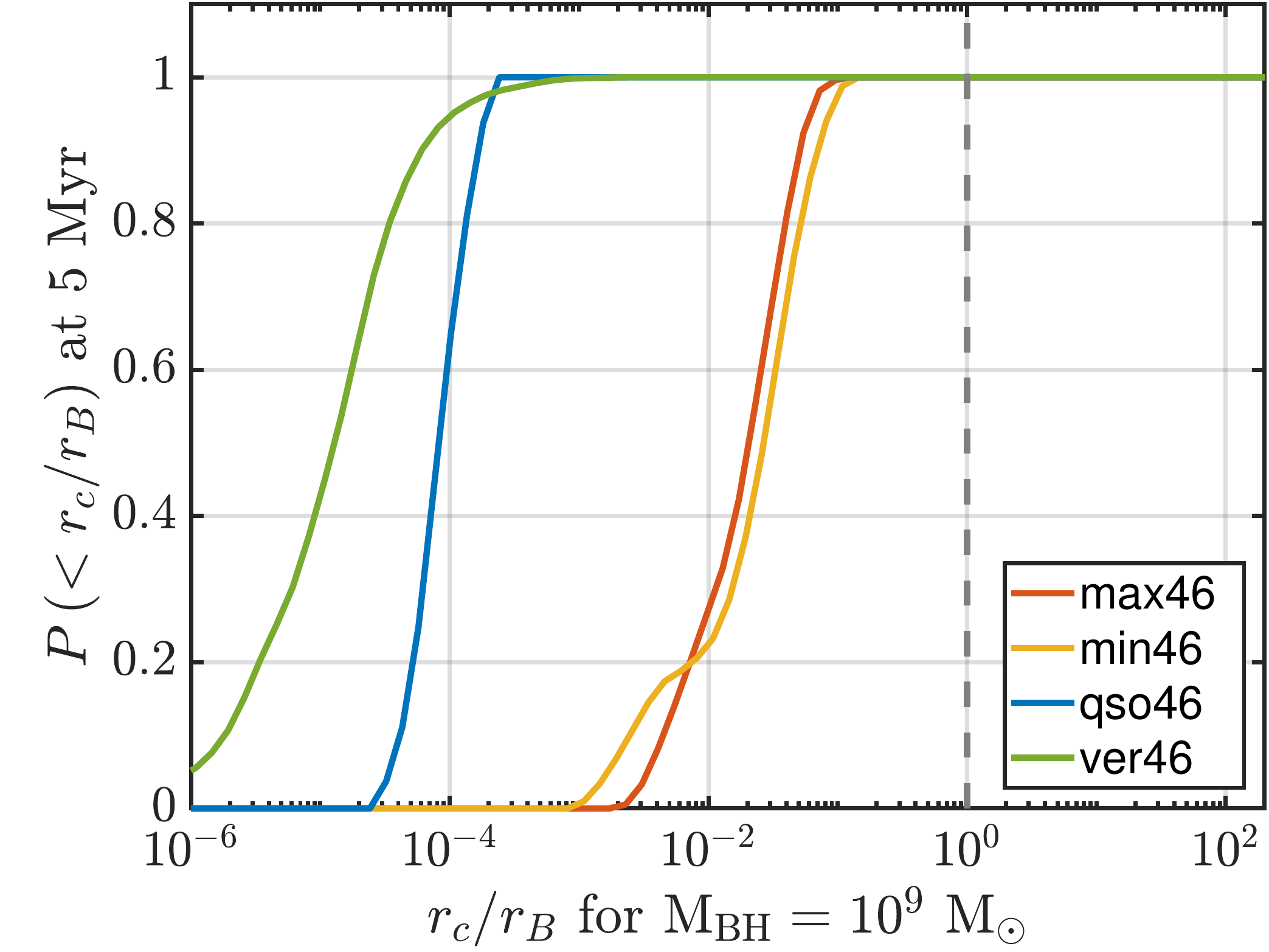}
\caption{Under the hypothesis that inflows interact only with $10^9$~M$_\odot$ central black hole, we express the circular radius $r_c$ in units of the Bondi radius. We plot the cumulative probability distribution of the ratio $r_\mathrm{c}/r_\mathrm{B}$ for the gas within $24$~pc. A 5-point moving-average smoothing is applied for visual clarity. Note that in the presence of cooling and gravity the Bondi radius would be much larger; the results in this figure are extreme upper limits to the ratio of $r_\mathrm{c}/r_\mathrm{B}$.}\label{fig:bondi}
\end{figure}

The cumulative probability distribution at $5$~Myr of  the ratio $r_\mathrm{c} / r_\mathrm{B}=({l c_s}/{(2GM_{\rm BH})})^2$ is shown in Figure \ref{fig:bondi}. We consider gas within $24$~pc, as in the left column of Figure \ref{fig:inflowmass}, assuming again a $10^9$~M$_\odot$ black hole. The qso runs have by far the smallest values of $r_B$, as they have much smaller $c_s$. Thus, with respect to Figure \ref{fig:orbit}, they move toward rather lower values than the jets.
At both low and high power most of the inflows circularize well within the Bondi radius, with the only exception of run min43, for which the distribution is significantly shifted to larger radii, and only 80 per cent of the gas penetrates within $r_\mathrm{B}$.  $r_\mathrm{c} / r_\mathrm{B}$ scales as $1/M_\mathrm{BH}^2$, so a BH with $M_\mathrm{BH}=10^8$~M$_\odot$ would be still have significant penetrating inflows ($>$90 per cent) for all low power runs except min43, and at high power for runs qso46 and ver46, while for min46 and max46 only 20 per cent of the inflowing gas would circularize within the Bondi radius.

The presence of radiative cooling in the simulation would result in lower values for $r_\mathrm{B}$, however in the p46 series the cooling time of the hot gas exceeds $10$~Myr or sometimes even the $100$~Myr, so they are the most accurate runs in this respect. Towards the end of the simulation ($15$~Myr or later), most of the gas around the BH has had the time to cool by adiabatic expansion, yielding lower values for $c_s$. So long-surviving inflows (as we observe in Figure \ref{fig:inflowmass}) tend to penetrate deeper within $r_\mathrm{B}$.

\section{AGN Feedback on the cold ISM}\label{sec:fb} 

\subsection{AGN-clump interactions}\label{sub:clump} 

\begin{figure*}
\includegraphics[width=1.\textwidth]{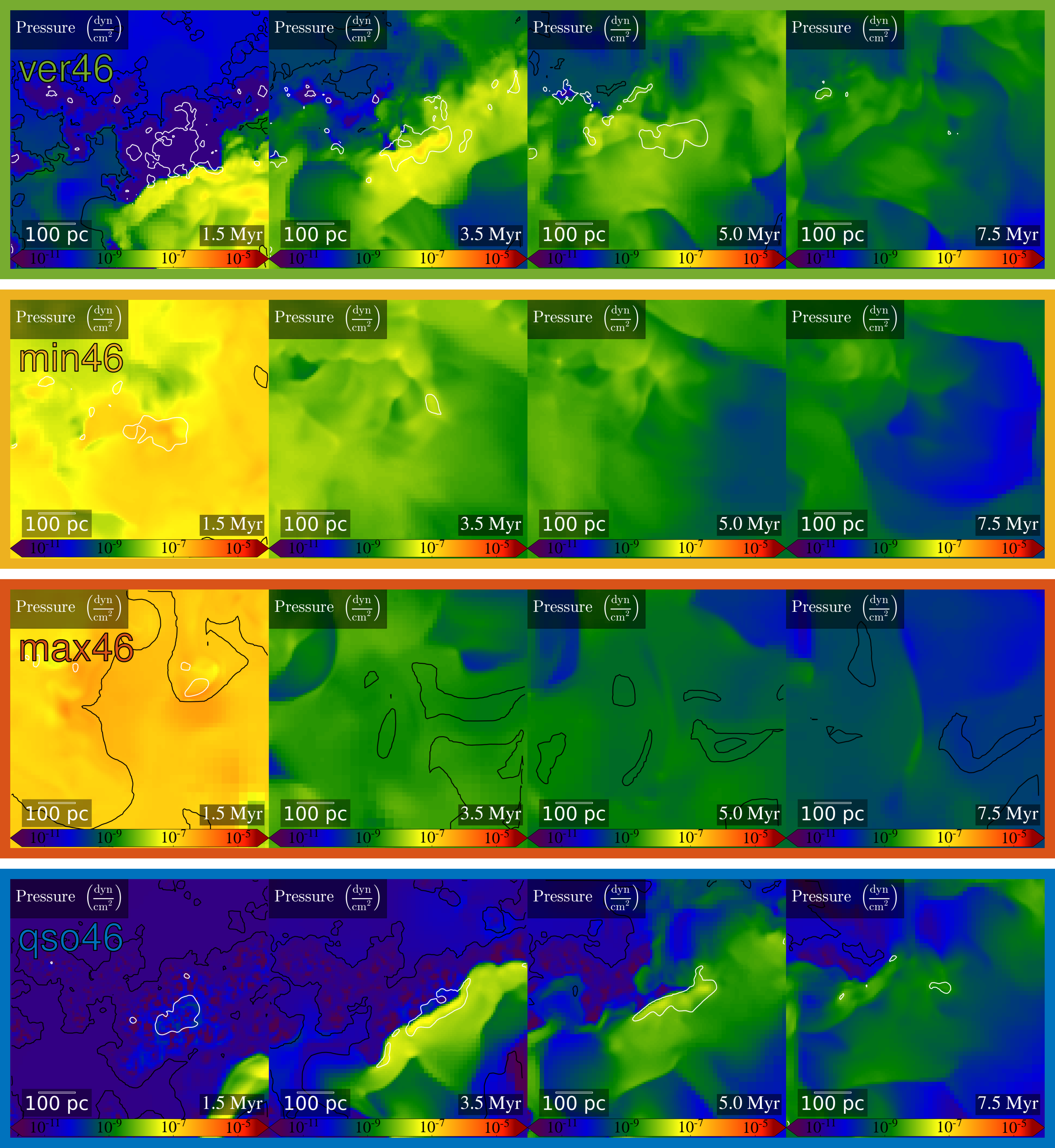}
\caption{Pressure evolution (view aligned with the z-axis) in the selected clump for the indicated runs. View: $500$~pc $\times500$~pc.The contours trace isodensity surfaces at $4$ (black) and $2500$~cm$^{-3}$ (white).  In run ver46 the densest clump core sees an increased pressure as the blastwave hits. In run min46 the clump is totally engulfed and pressurized; in max46 the clump is hit head-on and completely consumed by the jet beam.
In run qso46, the clump core was already heated and partially dispersed by the radiation. The cores have time and space to expand, thus the pressure enhancement is minimum. 
}\label{fig:clump46}
\end{figure*}

Figure \ref{fig:clump46} shows the evolution of the gas pressure in a $500$~pc region around a selected clump, located about $1$~kpc from the origin, roughly along the direction of the jet in the max runs (see also Figure \ref{fig:IC}). The clump is seemingly destroyed, sooner or later, by all high power AGN, though in different ways.

\begin{figure*}
\includegraphics[width=1.\textwidth]{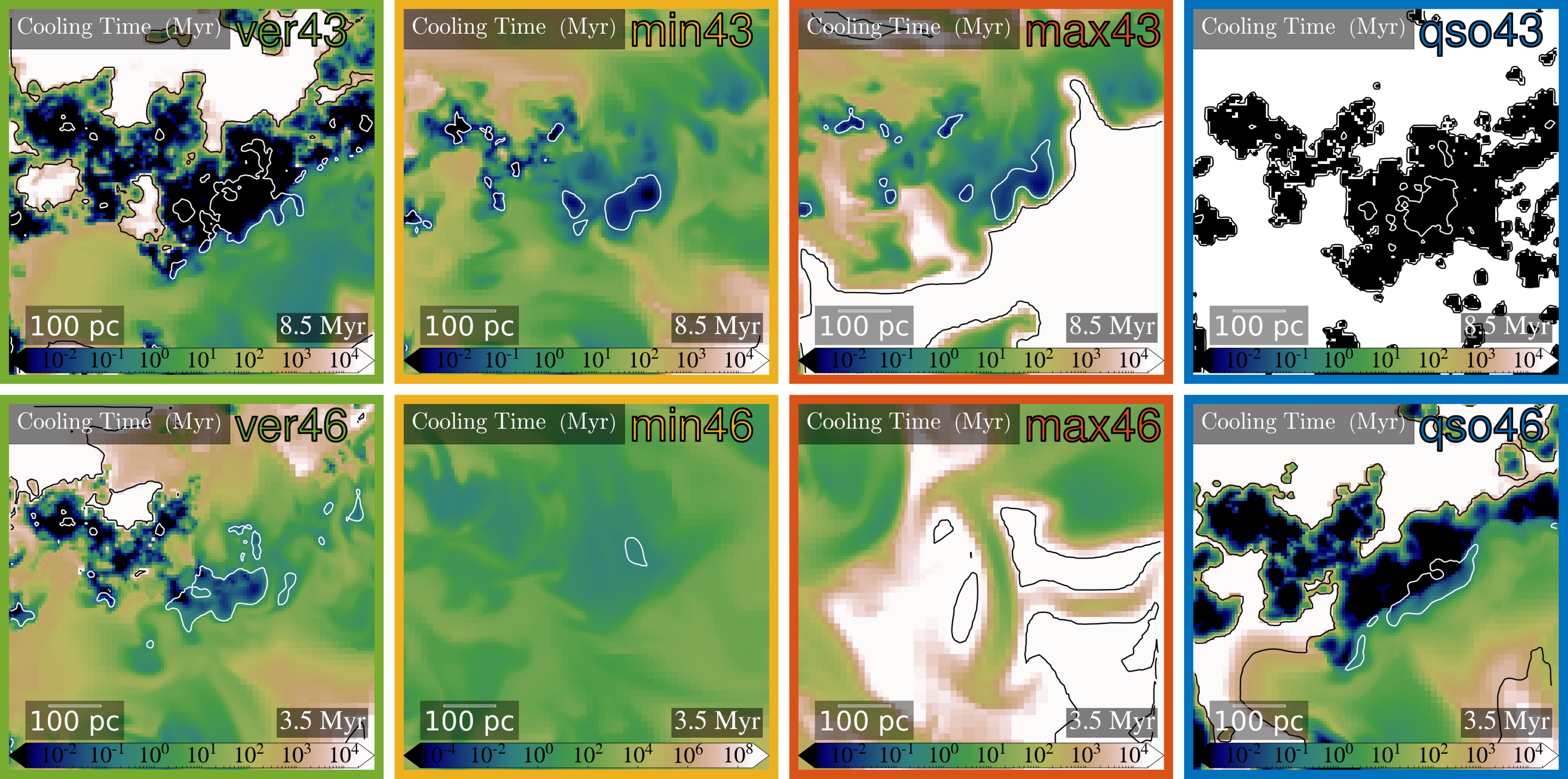}
\caption{Post-processing cooling time estimates, including molecular and atomic cooling, with the proper simulation metallicity, after \citet{vasiliev_cooling_2013}, as well as bremsstrahlung and pair production cooling from \citet{krause_multiphasejets_2007}. Same view and contours as Figure \ref{fig:clump46}, at $8.5$ (p43 series, top row), and $3.5$~Myr (p46 series, bottom row). Surrounding the clump with hot gas as the min/max runs do, is what lowers cooling rate the most, despite the pressure enhancement. A jet/clump direct collision is important only at high power, in max46. 
}\label{fig:clumpcool}
\end{figure*}

In run ver46  a shock is driven into the clump in the disc plane ($1.5$~Myr), which compresses it ($3.5$~Myr) and heats it. By the time the shock has passed ($5$ and $7.5$~Myr snapshots) very little dense gas is left (white contours), so the clump is effectively evaporated, but in a considerably longer time than  $t_{on}$. The shock driven into the clump in run qso46 is weaker, but the clump had been already partially heated by the IR radiation, as shown by the much shallower density contours at $1.5$~Myr. The most effective way of heating the clump is  shown by the min46 and max46 runs, in which the clump is completely engulfed in the shocked gas percolating through the whole disc, so the pressure all around the clump is very high, and most or all the structure is evaporated within $3.5$~Myr already. In addition, in run max46 the clump is located along the jet beam, so ram-pressure stripping and hydrodynamic instabilities play a larger role. For the low power cases, the description of how the AGN interacts with the gas is qualitatively very similar, but only runs min43 and max43 manage to affect and smooth out all the density peaks inside the clump, thus erasing the initial fractal structure, similar to the p46 runs. 

As a result of this smearing and the other interactions described above, the cooling time of the gas changes. Figure \ref{fig:clumpcool} displays a post-processing estimate of the cooling time  $t_{cool}$  of the gas, including molecular and atomic cooling after \citet{vasiliev_cooling_2013}, and bremsstrahlung and pair production cooling from \citet{krause_multiphasejets_2007}.



We conclude that  jet feedback is unlikely to significantly alter the properties of the fractal ISM, except along or very close to the jet beam, as in the max46 case.  Rapid cooling of the material stripped from the clumps may increase density, mass rate and duration of the fast, dense nuclear outflows the jets power (see Section \ref{sec:outflow}).

Finally we notice that despite the initial shock compression of the clump, its $t_{cool}$ is always increased. In other words, we observe no \emph{positive feedback}, not even transient, at variation with previous numerical models (\citealt{gaibler_jet-induced_2012,tortora_agn_2009}).
While this may depend on the fractal density distribution we assume, this is more likely indicating  the shocks in our simulations are thermally very efficient. Weaker shocks from weaker AGN, or the inclusion of cooling, would lead to lower shock temperatures, and thus likely allow for brief enhancement of the local cooling rates \citep{wagner_theory_2016}.



\subsection{Feedback on the gaseous disc}\label{sub:disk} 
\begin{figure*}
  \includegraphics[width=\columnwidth]{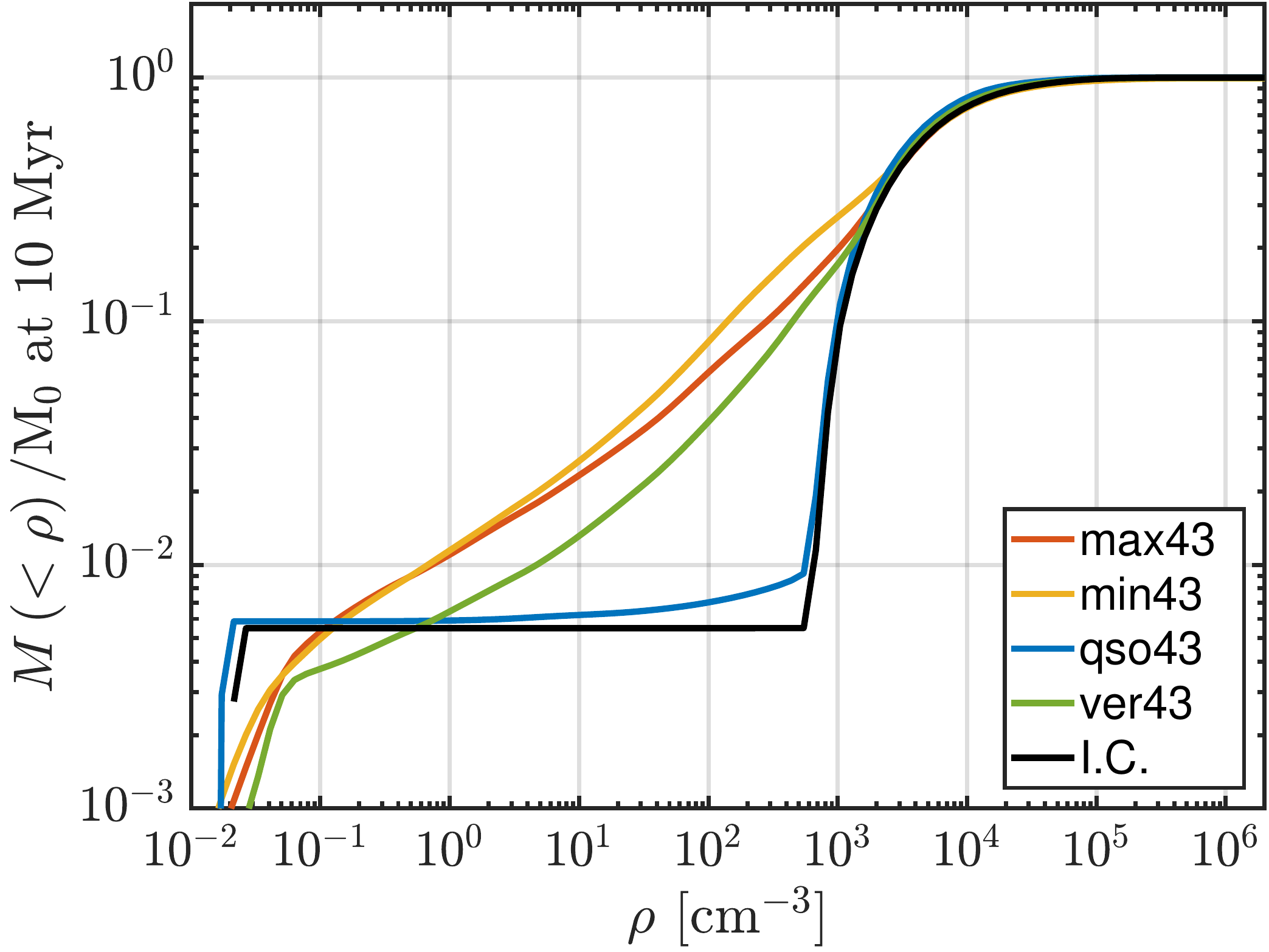}
  \includegraphics[width=\columnwidth]{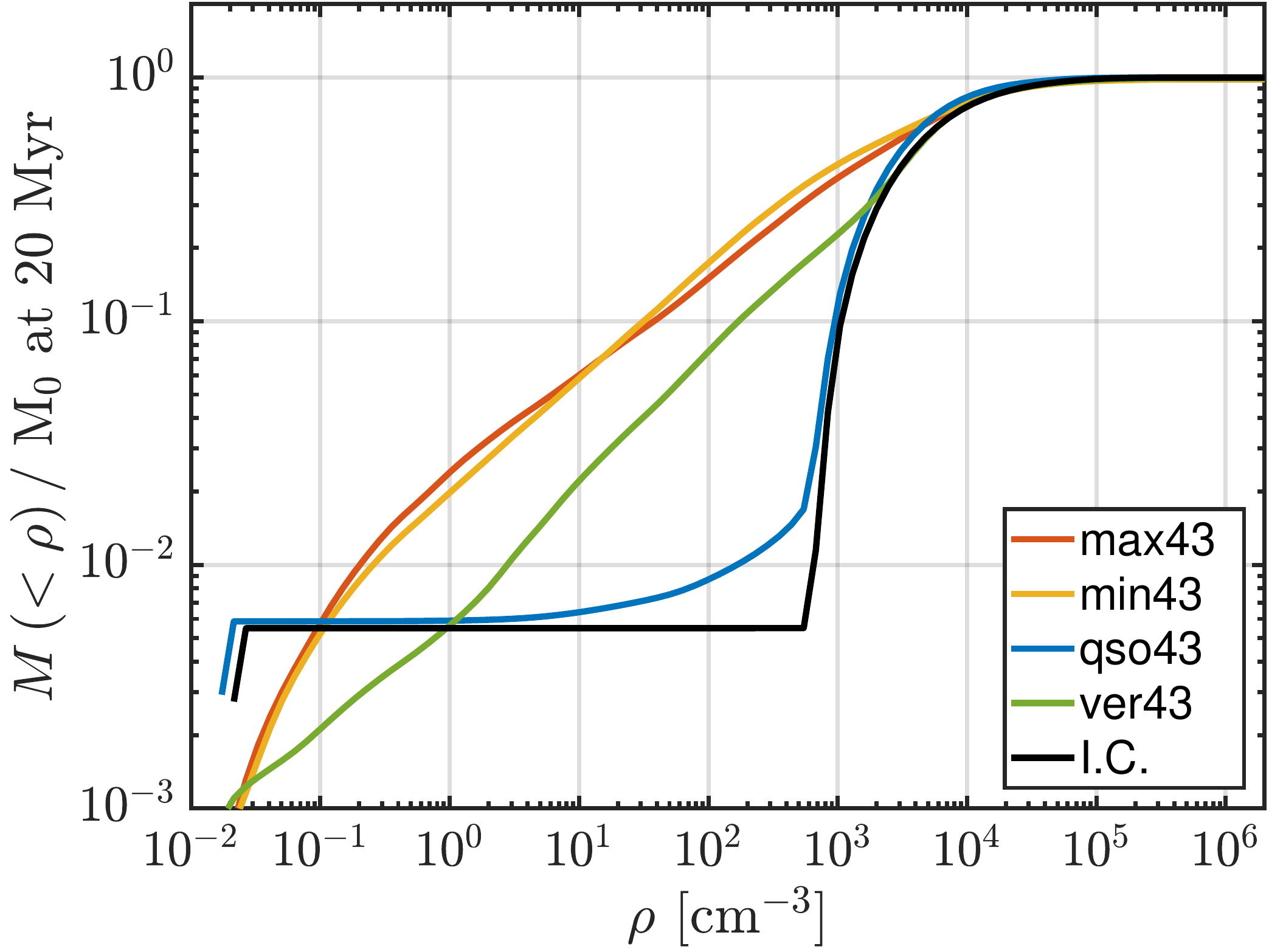}
  \includegraphics[width=\columnwidth]{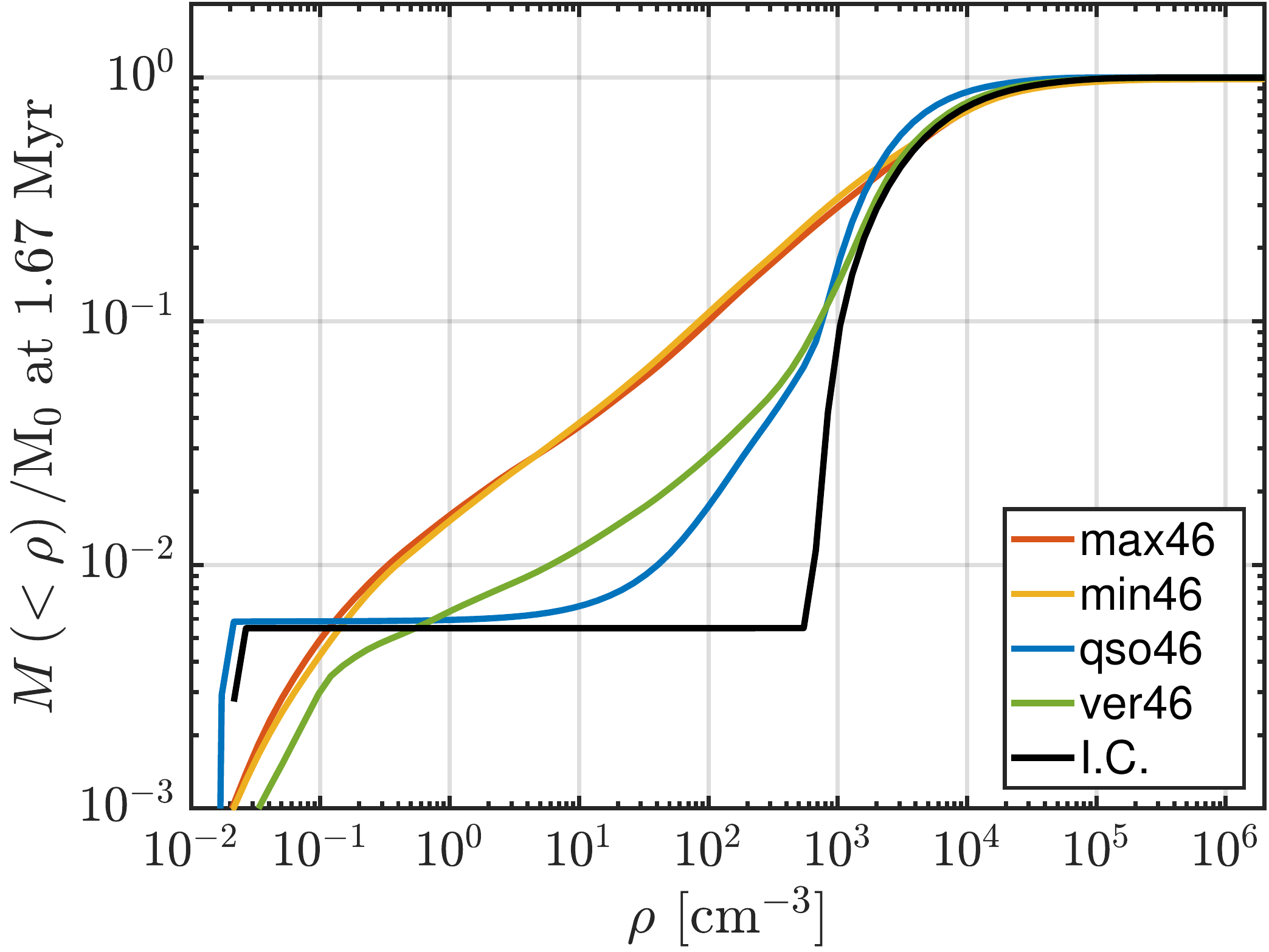}
  \includegraphics[width=\columnwidth]{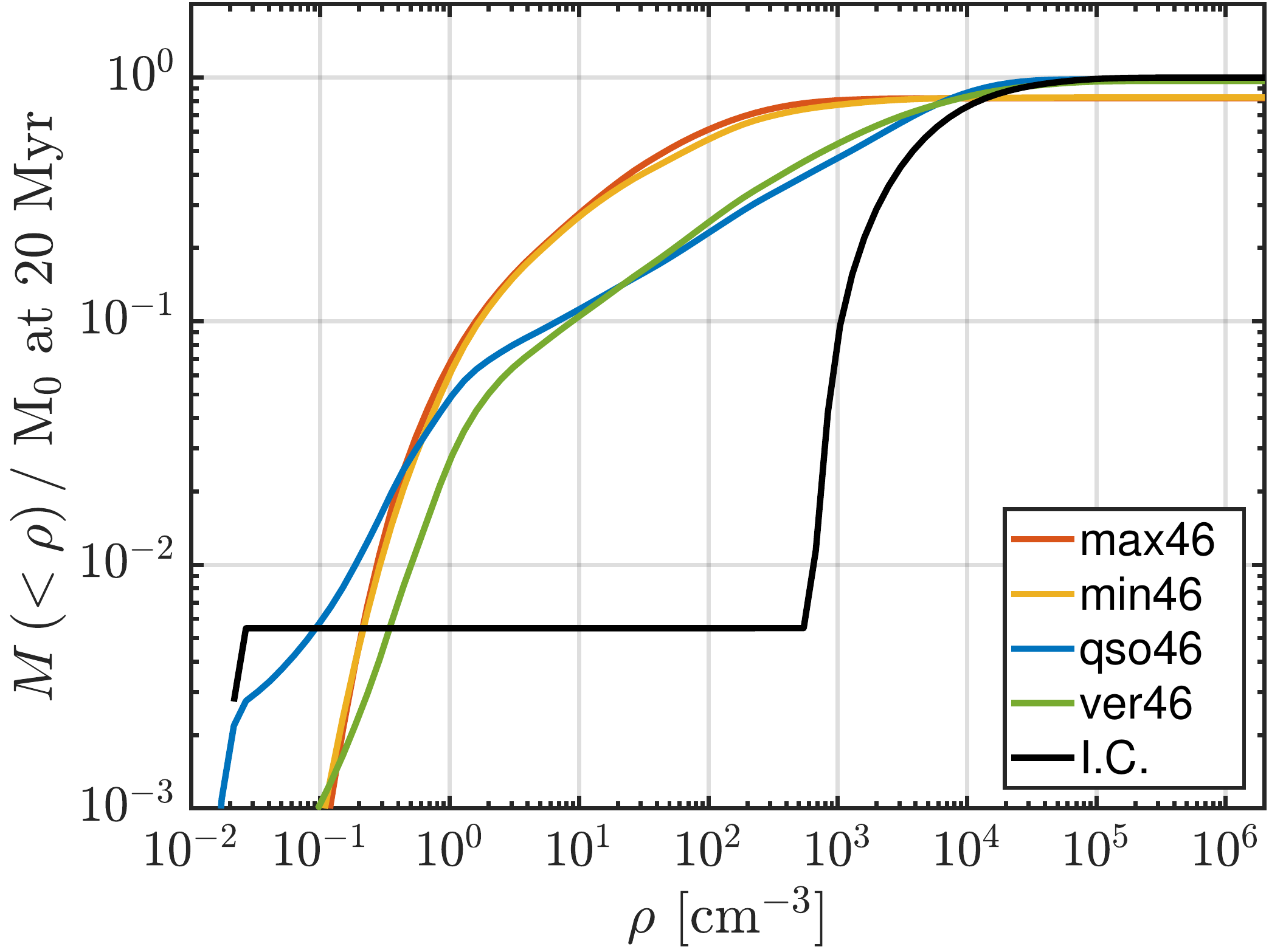}
  \caption{Cumulative density pdf for the p43 (top) and p46 runs (bottom), at time $t_{on}$ (left), and at the end of the simulation (right). The distributions are normalized at the initial total gas mass, $M_0 = 2.1\times10^{10}$~M$_\odot$. Runs min and max, at both powers are creating more intermediate gas density from the beginning; qso and ver (paired most of the time at high power) can do the same only at high power and at the end of the simulation. Run min46 and max46 are the only ones reducing appreciably the total gas mass, progressively removing the densest $20\%$ of the gas.
}\label{fig:denscumpdf}
\end{figure*}

Finally, we want to estimate the feedback effect on the global gas content of the disc and its star formation properties. In Figure \ref{fig:denscumpdf} we show the cumulative density probability distribution, at the time $t_\mathrm{on}$ and at the end of the simulation.

We include all gas inside a sphere of the same radius of the disc, in order to include the warm gas phase lifted off the disc by the AGN. The  probability is normalised to the initial mass, $M_0\sim2.15\times10^{10}$~M$_\odot$ included in the initial selection (recall that the total cold phase in the disc amounts to $2.1\times10^{10}$~M$_\odot$). 
The initial conditions of the disc are shown in black in all panels for comparison.
In the p43 runs, all jets heat and rarefy some dense gas, which populates the region of the diagram between $0.01$ and $100$~cm$^{-3}$. As time elapses, heating by the vertical jet or the radiation becomes less efficient compared to max43 and min43, and fails to affect the densest gas peaks. In this respect qso43 and ver43 differ little from the I.C., thus they having much less impact on the densest starforming regions than min43 and max43 simulations.

The high power set of simulations exhibit large difference betweeen min/max 46 on one side, and ver/qso 46 on the other. The first two have a similar but faster evolution than their low-power counterparts, while the ver46 jet leaves the halo too early to continue heating the gas. By the end of the simulation, it has affected only marginally more gas than ver43. As for the other high power runs at $20$~Myr, qso46 is the only one to show some compression of dense gas indicated by the peak around $1000$~cm$^{-3}$, which may lead to local burst of star formation (although compensated by a reduction at higher densities, so globally the gas is rarefied). 
Run max46 and min46 are the only ones to significantly reduce the total disc gas content, as they saturate to a much lower value (around $20\%$ lower). This effect is seen to be progressing over the course of the whole $20$~Myr and the gas content was still decreasing by the end of the simulation.


\section{Conclusions}\label{sec:conclusion} 
We have performed  HD and RHD simulations to compare the effects
of AGN feedback by jets and radiation in idealized galaxies, at fixed AGN power ($10^{43}$ and $10^{46}$~erg/s) and lifetime ($10$ and $1.67$~Myr, respectively) in a multi-phase ISM. In the radiative simulations, we modeled the emission, absorption, and propagation of photons and their interaction with the gas via
momentum transfer, photo-ionisation, and absorption/scattering on dust.   Jet feedback is modeled as a hydrodynamic source term of a straight, self-collimated beam. We studied three jet models: one of a jet perpendicular to the disk (ver model) and two others of jets lying in the disk plane, along the directions of maximum (max model) and minimum (min model) column density. We summarize our results in the following.

\subsection{Outflow driving and mechanical coupling}
\begin{itemize}
\item {\bf Hot outflows escape perpendicular to the disc.} Jet plasma and radiation are always preferentially ejected perpendicular to the disk. In the case of inclined jets, they get their beams deflected outside the disk plane. Denser outflows can also occur inside the disc plane. 

\item {\bf Momentum coupling is more effective for jets than radiation-driven winds}  Jets show very high momentum coupling, even a factor 100 (at high power) or 1000 (low power) with respect to the input momentum flux, as long as the jet is on and the jet streams are directly hitting dense gas. The coupling between the photons and the gas  is not as efficient, mainly because the heated and accelerated gas is confined in the central cavity and does not interact with large gas volumes.

\item {\bf Young jets power fast nuclear outflows.} At early times, the young jets can briefly accelerate small patches of dense gas ($\sim 100$~cm$^{-3}$) up to $1000$~km/s, more when in the disk plane; the qso runs never exceed a few $100$~km/s at those densities.

\item {\bf  Outflow properties converge with time} After $15$ or $20$ Myr, all AGN of the same power converge in mass flux and appearance of the outflows. An exception is the outflow in qso43, which remains confined in its initial overdensity.

\end{itemize}

\subsection{Inflows and consequences for accretion and duty cycles}
\begin{itemize}
\item {\bf Inflows follow outflows.}  AGN trigger secondary gas inflows spanning from $<25$ to $500$~pc, and they do so  by reverse shocks, ``bouncing'' pressure waves and backflows, which are a direct consequence of outflows. Such inflows entrain gas from the disc and last for several Myr, even after the source is switched off. 

\item {\bf  Feeding/feedback and duty cycle.} These low-to-moderate level inflows can easily reach and sustain $10^{-3}-10^{-2}$~M$_\odot$/yr. Most of this gas has very radial trajectories, can reach the SMBH and provide fuel for a new AGN episode on short timescales. Gas with more angular momentum circularizes within a few Bondi radii. Additional dynamical processes, e.g., gas condensation and free-fall, acting on longer timescales, could perturb this gas to maintain and modulate BH growth and AGN activity.
\end{itemize}

\subsection{Feedback onto the cold ISM/star formation}
\begin{itemize}
\item {\bf  Inclined jets percolate through the entire disc.}  Non-vertical jets, despite being deflected from the disc plane, always percolate very efficiently through the cold disc, heating and pressurizing the gas at much larger radii than vertical jets and radiation-driven winds.

\item {\bf  Inclined jets reduce star-forming gas} Runs min46 and max 46 remove the densest $20\%$ of the cold ISM by heating the clump cores, creating large amounts of warm gas with densities in the range $10-200$~H/cc, and potentially reducing the global SF significantly. However, gas cooling may counteract this process in the densest gas clumps, since the cooling time is never longer than $0.1$~Myr, even if AGN feedback is included.

\item{\bf  Three ways to heat a gas clump}. Three different behaviours emerged in the AGN/clump interaction for a specific large clump analyzed in more detail: in the ver runs the clump is hit by a shock front, which compresses and heats it from one side. The jet plasma in the min/max runs engulf the clump completely. In both cases the clump pressure is raised, and the highest density peaks reduced. In run qso46 by the time the blastwave hits the clump, the latter has already been heated by IR radiation, showing much shallower density contours. However, taking radiative cooling into account, the densest clumps are likely to cool efficiently and preserve their density peaks.

\item {\bf  Quenching star formation $\neq$ quenching SMBH accretion.} The runs that affect most of the cold gas content over the whole disc, min46 and max46, show inflow rates equal to if not higher than the other AGN of the same power, hinting that the AGN feedback can affect the two processes independently from each other.

\end{itemize}

This work is a first step towards a quantitative, controlled comparison of  AGN feedback from jets and radiation in gas-rich, multi-phase environments common to high-redshift, evolving galaxies. In order to understand galaxy formation in the presence of supermassive black holes, inclusion of both types of feedback, using realistic and detailed implementations as those presented here, is essential.

\section*{Acknowledgments}
%

We are grateful to Ramesh Narayan for suggesting to look at the angular momentum of the inflowing gas.

MV and SC acknowledge funding from the European Research Council under the European Community's Seventh Framework Programme (FP7/2007-2013 Grant Agreement no.\ 614199, project ``BLACK''). 

This work was granted access to the HPC resources of CINES under the allocations c2015047421 and x2015046955 made by GENCI. Part of the simulations have been performed on the draco cluster hosted by the Max Planck Computing and Data Facility (\url{http://www.mpcdf.mpg.de/}). 
Part of the analysis has made use of the Horizon cluster, hosted by the Institut d’Astrophysique de Paris. We warmly thank S. Rouberol for running it smoothly.

\vspace{-0.5cm}

\bibliographystyle{mn2e}
\bibliography{all}

\end{document}